\PassOptionsToPackage{prologue,dvipsnames}{xcolor}

\documentclass[acmsmall,screen]{acmart}

\newcommand{\infra}{\textsf{PanicFI}\xspace}
\newcommand{\dataset}{\textsf{Panic4R}\xspace}
\newcommand{\tool}{\textsf{PanicKiller}\xspace}
\usepackage{makecell}
\usepackage{listings} % 展示代码
\usepackage{multirow}
\usepackage{enumitem}
\usepackage{threeparttable}
\usepackage{rotating}
\usepackage{hyperref}
\usepackage[ruled,linesnumbered,vlined]{algorithm2e}
\usepackage{graphicx}
\usepackage{xcolor}
\definecolor{bg}{rgb}{0.95,0.95,0.95}
\usepackage{caption}
\usepackage{ragged2e}
\usepackage{booktabs}
\usepackage{subfigure}
\usepackage{pifont}
\usepackage{url}
\usepackage{amsmath}
\usepackage{colortbl}
\usepackage{wrapfig}
\usepackage{framed}
\AtBeginDocument{%
  }

\setcopyright{acmlicensed}
\copyrightyear{2024}
\acmYear{2024}
\acmDOI{XXXXXXX.XXXXXXX}

\acmISBN{978-1-4503-XXXX-X/18/06}

\begin{document}

% \title{Towards Fixing Panic Bugs for Real-world Rust Programs}

\title[PanicFI: An Infrastructure for Fixing Panic Bugs]{PanicFI: An Infrastructure for Fixing Panic Bugs in Real-World Rust Programs}

\author{Yunbo Ni}
\email{yunboni@smail.nju.edu.cn}
\affiliation{%
  \institution{State Key Laboratory for Novel Software Technology \\Nanjing University}
  \city{Nanjing}
  \country{China}
}

\author{Yang Feng}
\email{fengyang@nju.edu.cn}
\affiliation{%
  \institution{State Key Laboratory for Novel Software Technology \\Nanjing University}
  \city{Nanjing}
  \country{China}
}

\author{Zixi Liu}
\email{zxliu@smail.nju.edu.cn}
\affiliation{%
  \institution{State Key Laboratory for Novel Software Technology \\Nanjing University}
  \city{Nanjing}
  \country{China}
}

\author{Runtao Chen}
\email{211220018@smail.nju.edu.cn}
\affiliation{%
  \institution{State Key Laboratory for Novel Software Technology \\Nanjing University}
  \city{Nanjing}
  \country{China}
}

\author{Baowen Xu}
\email{bwxu@nju.edu.cn}
\affiliation{%
  \institution{State Key Laboratory for Novel Software Technology \\Nanjing University}
  \city{Nanjing}
  \country{China}
}

\begin{abstract}
The Rust programming language has garnered significant attention due to its robust safety features and memory management capabilities.
Despite its guaranteed memory safety, Rust programs suffer from runtime errors that are unmanageable, i.e., panic errors.
Notably, traditional memory issues such as null pointer dereferences, which are prevalent in other languages, are less likely to be triggered in Rust due to its strict ownership rules. 
However, the unique nature of Rust's panic bugs, which arise from the language's stringent safety and ownership paradigms, presents a distinct challenge. 
Over half of the bugs in rustc, Rust’s own compiler, are attributable to crash stemming from panic errors.
However, addressing Rust panic bugs is challenging and requires significant effort, as existing fix patterns are not directly applicable due to the design and feature of Rust language. 
Therefore, developing foundational infrastructure, including datasets, fixing patterns, and automated repair tools, is both critical and urgent.

% However, understanding root causes and resolving these panics often requires substantial effort due to the limited information provided, and the stacktrace could be intricate, often omitting the actual fault locations.
% Although numerous automated program repair techniques exist, we observe that the prevailing fix patterns do not readily apply to Rust programs due to natural differences in language mechanisms.

This paper introduces a comprehensive infrastructure, namely \infra, aimed at providing supports for understanding Rust panic bugs and developing automated techniques.
In \infra, we construct a dataset, \dataset, comprising 102 real panic bugs and their fixes from the top 500 most-downloaded open-source crates. 
Then, through an analysis of the Rust compiler implementation , we identify Rust-specific patterns for fixing panic bugs, providing insights and guidance for generating patches. 
Moreover, we develop \tool, the first automated tool for fixing Rust panic bugs, which has already contributed to the resolution of 28 panic bugs in open-source projects. 
The practicality and efficiency of \tool confirm the effectiveness of the patterns mined within \infra. Furthermore, \dataset serves as a benchmark for evaluating APR tools focused on Rust panic bugs. 
We believe the construction and release of \infra could enable the expansion of automated repair research tailored specifically to Rust programs, addressing unique challenges and contributing significantly to advancements in this field.

% effectively generates correct patches on the real-world large-scale dataset, and has already assisted in the resolution of 28 panic bugs in open-source projects. Each resolved issue has been validated by their developers and merged into the respective codebases.
\end{abstract}


\maketitle

\section{Introduction}

As a statically typed programming language, Rust has gained popularity for its well-known memory safety guarantees and high performance.
Recently, the White House Office of the National Cyber Director also emphasized the necessity of using programming languages that have fewer memory safety vulnerabilities~\cite{WhiteHou52:online}, and nominating Rust as an example of a memory-safe programming language. 
The foundational principles of Rust, including ownership, borrowing, and lifetimes, enable developers to implement secure and efficient programs. 
Rust's emphasis on zero-cost abstractions and fearless concurrency has significantly contributed to its popularity in systems programming~\cite{bugden2022rust, jung2021safe, klabnik2023rust, jung2020understanding}.
This design has led to an increase in the development of widely recognized software projects written in Rust~\cite{Servothe7:online, RedoxYou24:online, TiKVTiKV64:online, StratisS70:online, Cloudfla15:online}.

Although Rust boasts security features and significantly reduce common bugs such as null pointer dereference, uninitialized variables, and data races, it could suffer from \textit{panic errors}, which are caused by a Rust-specific error-handling mechanism.
The severity of the panic is evident from its typical consequences—program crashes or terminations that may lead to the improper handling of resources, such as unclosed file descriptors or network connections~\cite{Zheng2023A}. 
Unlike Java's structured exception handling framework that manages routine errors, Rust's panic mechanism is designed for unrecoverable situations, significantly impacting program stability. 
% Panics can occur due to various reasons, such as failed assertions, index out-of-bounds errors, or attempts to access invalid memory~\cite{Astrauskas2020how}. 
Moreover, the Rust compiler, rustc, written in Rust, also exhibits vulnerabilities to panic bugs. 
\textit{Over half} of the issues in Rust language's official GitHub repository are categorized as Internal Compiler Errors (ICE)~\cite{Issues·r78:online}, primarily caused by panic bugs in rustc. 
Therefore, understanding and resolving panic bugs in Rust is crucial for ensuring the reliability and stability of Rust programs.

% However, fixing panic bugs can be a tedious and challenging task for Rust developers.
However, the infrastructure and toolchain supporting the Rust language are not yet as mature as those for other languages. 
Most existing code fix datasets and automated program repair (APR) tools are designed for languages like Java and C/C++. 
These tools can be challenging to adapt to Rust due to significant differences in language mechanisms, which may lead to violations of ownership rules or even failing to compile. 
For instance, commonly used strategies for fixing null pointer errors in Java are entirely inapplicable in Rust, as Rust's language design inherently disallows null values. Similarly, repair patterns for memory and pointer errors in C/C++ are not transferrable to handling panic-related bugs in Rust, as Rust's safety guarantees prevent such memory errors from occurring.
In addition, Rust’s unique memory management model and lifetime rules leads to a steep learning curve, highlighting the urgent need for infrastructure to support panic bug fixes, particularly in complex, large-scale and real-world Rust programs.

% On the one hand, the stack expansion information output by the compiler after a panic error is triggered could be complex and hard to understand.
% Regarding dozens of lines of stack stacktrace, it is challenging to understand the root cause, localize the bug, and fix the program.
% Due to the unique memory management model and lifetime rules of Rust, the learning curve becomes particularly steep~\cite{Crichton2021The, crichton2024profiling, Qin2020Understanding}.

Due to the lack of mature infrastructure, such as datasets and repair patterns, fixing panic bugs can be a tedious and challenging task for Rust developers.
Recently, a few program repairing tools have been proposed for Rust, yet they are insufficient to address the most severe panic-related bugs.
Rust-lancet~\cite{yang2024lancet} was developed to tackle bugs related to violations of ownership rules through three specific strategies. 
However, these strategies are tailored exclusively to Rust's ownership rules, which are verified prior to runtime, making them unsuitable for addressing panic bugs. 
Similarly, Ruxanne~\cite{robati2024patterns} and RustAssistant~\cite{deligiannis2023rustassistant} focus on common compilation issues, such as incorrect data types, but these patterns do not effectively mitigate panic bugs.

To overcome the aforementioned challenges and fill the program understanding gap, in this paper, we design and implement the first infrastructure \infra aiming at automatically fixing panic bugs for real-world Rust programs.
We have constructed a dataset \dataset, containing more than $100$ panic bug instances and corresponding fix patches, derived from open source projects in the ecosystem.
% Specifically, we constructed $52$ real fix patches by extracting the pull requests for fixing panic bugs in the top 100 downloaded open-source projects of Rust crates.
% Besides, we also simulated real-world development and manually constructed $40$ pairs of panic-triggering code and fix patches to guarantee the diversity of panic root causes in the dataset.
Referring to the Rust implementation code, we further perform fix pattern mining with Rust-specific syntactic features to provide a reference for the community to better understand panic bugs and fixing strategies.
Further, we implement an automated fixing tool for panic bugs, namely \tool, which first applies dependency analysis for cross-file level error localization, then combines it with semantic information of the reported errors for fix pattern matching, and finally outputs sorted patches with scores and descriptions.

To evaluate the effectiveness of \tool, we conducted extensive experiments on \dataset. 
For fault localization, \tool has achieved high accuracy at different granularities.
We also compared \tool to the LLM-based ChatGPT-4.0, employing both single and multi-round conversations as baselines. 
Results demonstrate that \tool outperforms ChatGPT-4.0 in fault localization and patch generation, highlighting its practicality and reliability. Moreover, \tool has effectively resolved issues in open-source Rust projects, with 28 panic bug fixes validated and merged by developers.
In summary, the contributions of this work are as follows:
\begin{itemize}
    % \item \textbf{Infrastructure.}
    % We proposed a comprehensive infrastructure, \infra, to support fixing panic bugs of real-world Rust programs. This infrastructure includes a dataset, mined fixing patterns, and an APR tool. \infra aims to bridge critical gaps in the research on Rust programs and program repair methods.

    \item \textbf{Dataset.} 
    We constructed the first public dataset for Rust panic bugs, named \dataset, which comprises 102 real bugs and their corresponding fixes from PR records of the top 500 most downloaded Rust crates. Each bug-fix pair is meticulously organized and has undergone thorough manual verification, facilitating future research.
    
    \item \textbf{Patterns.} 
    We mined a series of fix patterns for panic bugs compliant with Rust syntax. The potential application for these mined patterns could be developing automated repair tools, providing references for developers, serving as a dataset for fine-tuning LLMs, etc.

    \item \textbf{Tool.} 
    We introduced \tool, an automated repair tool specifically for Rust panic bugs, designed to address issues in real-world and large-scale Rust programs. Our experiment results show that \tool is more efficient than commercial LLM-based tools and has successfully resolved 28 open issues in Rust projects on GitHub.
\end{itemize}

% The dataset, mined patterns, and the source code of \tool can be found at \url{https://sites.google.com/view/panickiller/home}.

\section{Preliminary and Motivation}
% Rust, a statically typed programming language, is gaining popularity for its well-known memory safety guarantees and high performance, making it notably distinct from C/C++ and JVM family of languages.

% For Rust programs, the memory and thread safety is ensured through a specific mechanism known as the ownership system, which naturally prevents common errors like null pointer errors.
Rust language leverages ownership mechanism to manage memory safety and thus naturally prevent common errors like null pointer errors.
This language mechanism enforces strict rules at compile-time that manage memory usage and ownership, effectively eliminating common memory issues such as dangling pointers and data races. 
% However, while Rust's ownership model addresses many memory-related problems, Rust programs could still suffer from panic bugs, arising from scenarios such as array index out-of-bounds, which the ownership system does not inherently prevent. 
Rust language present panic mechanism for unrecoverable errors that signal bugs or critical conditions where continuing execution is not safe. 
This mechanism is different from the error handling approaches of many other languages, such as Java, C, and Python, which allow programs to catch errors and potentially recover from them.

\begin{wrapfigure}{r}{8.3cm}%靠文字内容的右侧
\centering
% \vspace{-15pt}
  \includegraphics[width=0.95\linewidth]{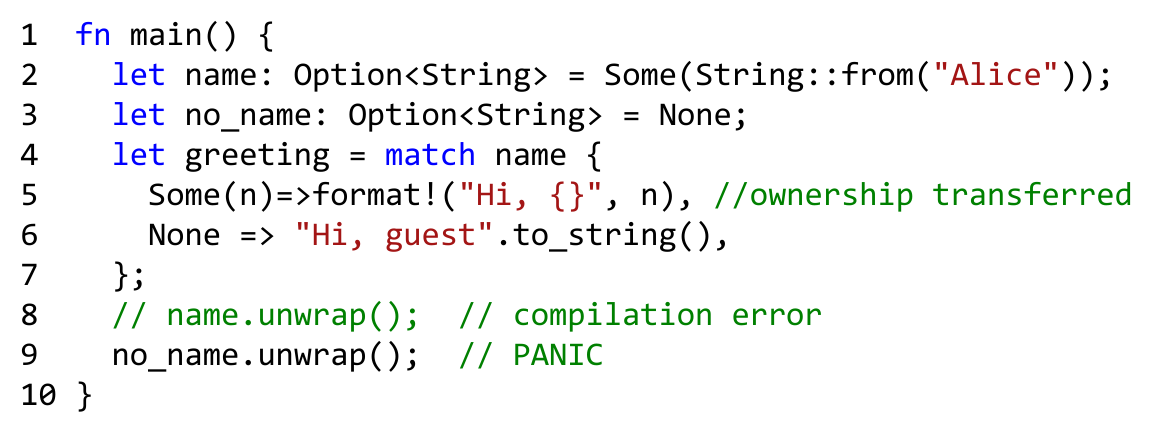}
  \vspace{-8pt}
  \caption{An example of Rust program.}
  \vspace{-8pt}
  \label{lst:motiv_rust}
\end{wrapfigure}

For example, consider the code snippet presented in Figure~\ref{lst:motiv_rust}, where line 2 declares a variable named \texttt{name} of type \texttt{Option<String>}. 
This type encapsulates a \texttt{String} object and employs \texttt{Option}, a Rust-specific data structure designed to handle potentially null values.
It uses \texttt{Some} to indicate the presence of a specific value and \texttt{None} to represent the absence of any value, as demonstrated in line 3.
In the subsequent \texttt{match} expression, the scenarios of \texttt{Some} and \texttt{None} for \texttt{name} are addressed separately. 
If \texttt{name} contains a value, this value is assigned to the variable \texttt{n}, which is then utilized to construct a new string. During this process, the ownership of \texttt{n} is transferred from \texttt{name}, rendering \texttt{name} subsequently null.
As a result, in line 5, the ownership of \texttt{name} has been transferred to \texttt{greeting} via the \texttt{match} expression. 
Consequently, \texttt{name} is now \texttt{None}. Attempting to unwrap it in line 8 will trigger a compile-time error due to the ownership transfer.
Also, as shown in line 9, calling \texttt{unwrap()} without checking the null value will result in a runtime error, i.e., \textit{panic error}, and the program terminates because it cannot handle the exception.

% \begin{listing}[h]
% \vspace{-10pt}
% \begin{minted}[linenos=false,
% %frame=single,
% fontsize=\scriptsize,
% highlightcolor=gray!10,
% highlightlines={5,8,9}]{rust}
% 1 fn main() {
% 2   let name: Option<String> = Some(String::from("Alice"));
% 3   let no_name: Option<String> = None;
% 4   let greeting = match name {
% 5     Some(n)=>format!("Hi, {}", n), //ownership transferred
% 6     None => "Hi, guest".to_string(),
% 7   };
% 8   // name.unwrap();  // compilation error
% 9   no_name.unwrap();  // PANIC
% 10 }
% \end{minted}
% \vspace{-10pt}
% \caption{An example of Rust program.} \label{lst:motiv_rust}
% \end{listing}
% %\vspace{-10pt}

\textbf{Motivation.}
The distinctive memory checking mechanism and stringent ownership transfer rules in Rust present significant challenges for understanding Rust programs and debugging them.
Besides, if a complex program encounters a panic bug at runtime, much effort is required to perform error localization and determine the root cause.
Although there have been many APR techniques practiced on Java/C/C++, due to the unique design mechanics of Rust, most of the existing repair patterns are not applicable to Rust programs.
Specifically, Java uses \texttt{null} to represent a variable that points to no object in memory. 
While convenient for indicating uninitialized states, this design often leads to problems, with most of Java's fix patterns addressing null pointer exceptions~\cite{liu2019tbar, koyuncu2019ifixr, kim2013par, liu2019avatar, durieux2017nullpointer}.

\begin{wrapfigure}{r}{7.5cm}%靠文字内容的右侧
\centering
% \vspace{-15pt}
  \includegraphics[width=0.95\linewidth]{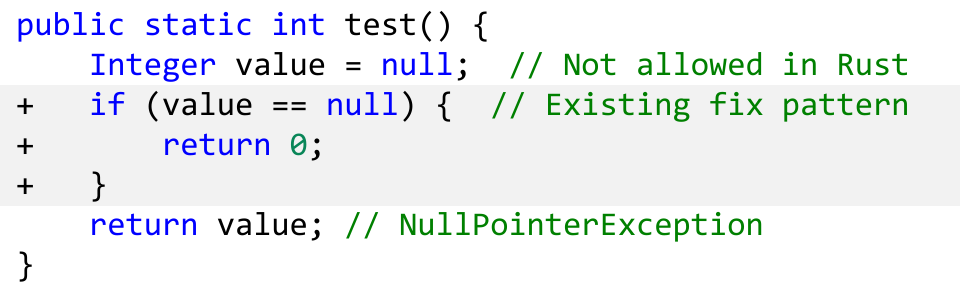}
  \vspace{-8pt}
  \caption{A minimized Java program triggering null pointer exception and the corresponding fix patch.}
  \vspace{-8pt}
  \label{lst:motiv_java}
\end{wrapfigure}
For example, the function in Listing~\ref{lst:motiv_java} tries to return a null value while the expected return type is \texttt{int}.
It would trigger a very common null pointer bug, and the existing APR tool generally adds an extra condition to check if the \texttt{value} is \texttt{null}, then return a random data that fits the type requirement.
% \begin{listing}[h]
% \begin{minted}[linenos=false,
% %frame=single,
% fontsize=\footnotesize,
% highlightcolor=gray!10,
% highlightlines={3,4,5}]{java}
% public static int test() {
%     Integer value = null;  // Not allowed in Rust
% +   if (value == null) {  // Exisiting fix pattern
% +       return 0;
% +   } 
%     return value; // NullPointerException
% }
% \end{minted}
% \vspace{-5pt}
% \caption{A minimized Java program triggering null pointer exception and the corresponding fix patch.} 
% \label{lst:motiv_java}
% %\vspace{-5mm}
% \end{listing}
However, conventional null pointer remediation strategies are largely ineffective in Rust due to the language's rigorous handling of null values. 
As demonstrated in Figure~\ref{lst:motiv_rust}, Rust does not employ a direct equivalent of \texttt{null}. 
Instead, it utilizes the \texttt{Option<T>} data type to explicitly manage scenarios where values might be absent. 
Data that could be \texttt{null} must be encapsulated within an \texttt{Option<T>}, effectively preventing null pointer errors inherent to Rust's design. 
Consequently, traditional fix patterns that address null pointer issues do not apply to Rust programs.
Besides, we notice that the C/C++'s bug fixing dataset, CVE-Fixes~\cite{fu2022vulrepair, chen2023vrrepair}, covers a wide variety of memory bugs such as improper restriction of operations within the bounds of a memory buffer (CWE-119), out-of-bounds write (CWE-787), and null pointer dereference (CWE-476), etc.
However, these specific bug types are inherently untriggerable in safe Rust programs due to the language's stringent safety guarantees. Thus, the methods commonly applied in C/C++ contexts are not transferable to Rust programs.

% In this paper, we concentrate on panic errors specific to Rust. We conduct a systematic study of such bugs by collecting data on common panic instances and their repair strategies from real-world and large-scale Rust projects. 

\begin{figure*}[h]
    \centering
    \includegraphics[width=0.97\linewidth]{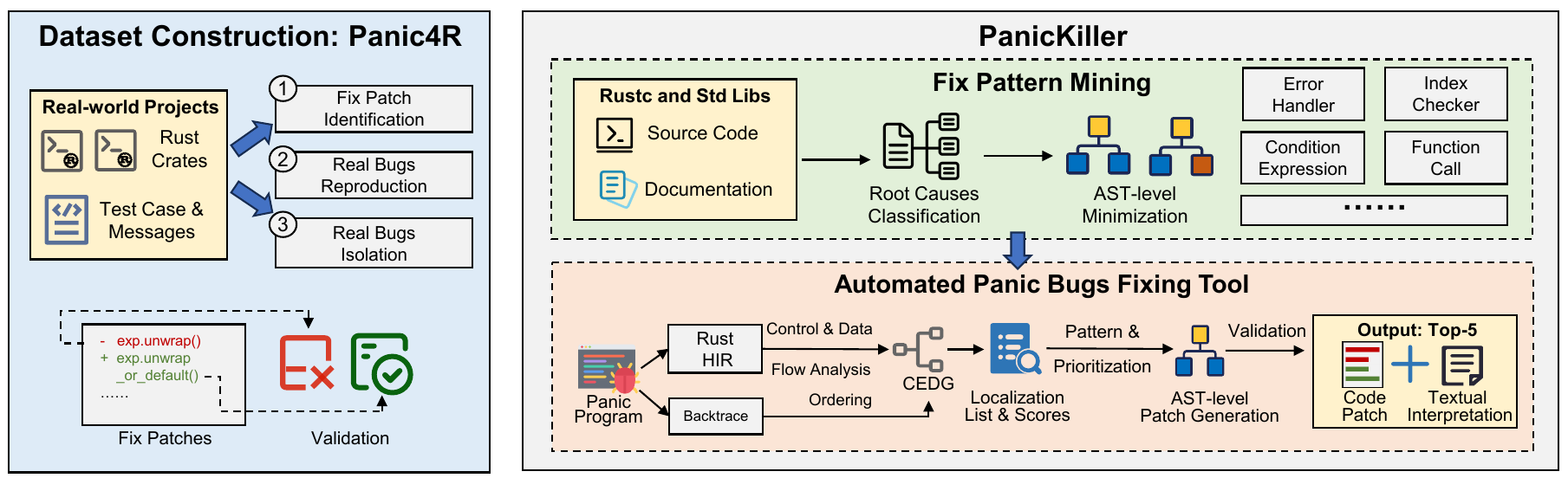}
    \vspace{-5pt}
    \caption{Key components of the proposed infrastructure \infra.}
    \label{fig:overview}
    \vspace{-10pt}
\end{figure*}

% \section{Infrastructure}
% To support research on understanding Rust program and automated program fixing, we construct a systematic infrastructure with dataset and testing tools.
% Specifically, 
\textbf{Our work.} To address the aforementioned challenges, we construct a systematic infrastructure infra with dataset, fixing patterns and testing tools, the key components of which are illustrated in Figure~\ref{fig:overview}.
Initially, we gather real-world panic bugs and their corresponding patches from the ecosystem to construct a dataset, \dataset, the specifics of which are detailed in Section~\ref{subsec:dataset}. 
Subsequently, referring to the implementation of Rust's compiler rustc, and its standard libraries, we identify panic-fix patterns, including abstract patches and natural language descriptions, as elaborated in Section~\ref{subsec:pattern_mining}. 
Finally, we develop an automated fixing tool, \tool, utilizing the extracted patterns. \tool conducts dependency analysis on source programs, organizes patch priorities, and generates hybrid fixing suggestions, further explained in Section~\ref{subsec:tool}.

\section{Dataset Construction} \label{subsec:dataset}

To build a dataset containing real-world panic bugs and their patches, we follow the construction process of Defects4J~\cite{just2014defect4j}, which is the most classic dataset in the APR task, and select the top $500$ downloaded crates from the public repository of Rust crates~\cite{cratesio6:online} for real bugs and fix patches extraction, so as to construct a real dataset.
Specifically, we manually employ the following workflow to collect code and thus guarantee the quality and reliability of the dataset.

\textit{(1) Identifying Real Fix Patches.}
For each target crate, we review the list of pull requests (PRs). If a closed PR's title or description contains keywords like ``fixing/repairing panic," it is considered a potentially valid fix patch. We also assess PRs linked to issues with keywords such as ``panic/thread panicked at...".
Then, we further examine the content of the code changes corresponding to PRs and their related PRs to ensure that they contain fixes for a panic bug.
For example, PRs that merely add or remove test cases without modifying the crate's source code are not considered relevant to our dataset. 
Similarly, changes to the API documentation for the crate are also outside the scope of our dataset construction. 
We only select closed PRs to ensure that they are either approved by developers or submitted by the developers themselves.

\textit{(2) Reproducing Real Bugs.}
To ensure the reproducibility of panic bugs in each crate, which requires specific versions, we download the source code for the corresponding commit IDs before and after the PR commit.
The two versions of the target crate are regarded as $V_{bug}$ and $V_{fix}$, respectively.
Then, we refer to the description in the PR or the code in the corresponding issue to minimize the test case $T_{test}$ that triggers the bug, i.e., calling the API of the target crate. 
We ensure that $T_{test}$ can trigger panic bugs and be compiled before and after switching versions, respectively.
% Additionally, both the buggy and the fixed program can pass regression test cases, which is exactly the same as our validation process in Sec~\ref{subsubsec:generation}.

\textit{(3) Isolating Real Bugs.}
In the target crate's repository, each PR commit may contain more than one code change, such as adding other functional logic or modifying data structures for new features, but not as a fix patch for panic.
To ensure the accuracy of the fix patches in the dataset, we manually verified code changes and kept only the patches used to fix the panic bug.
If a PR contains multiple fixes for panic bugs, we also split it into multiple code-fix pairs.
As a result, the patches for $V_{fix}$ versus $V_{bug}$ have no irrelevant content and are the precise changes necessary to fix the panic bug.
To summarize, each of the real datasets we build contains two crate versions, $V_{bug}$ and $V_{fix}$, and a test case $T_{test}$ that triggers panic bugs on $V_{bug}$ and is compilable on $V_{fix}$.
Table~\ref{tab:dataset} shows the details of the real-world dataset \dataset.

% In addition to the real-world dataset, we also manually constructed $40$ synthetic dataset to ensure diversity.
% The manually constructed datasets are formatted to precisely mirror the real datasets, with each program comprising 2-5 Rust files, each containing 10-30 lines of code.

\begin{table}[h]
\centering
\footnotesize
\caption{Statistics of bugs and patches available in \dataset.}
\vspace{-5pt}
\label{tab:dataset}
% \begin{threeparttable}
\setlength{\tabcolsep}{1mm}{
\begin{tabular}{ccccccc}
\toprule
\textbf{\begin{tabular}[c]{@{}c@{}}Download\\  Ranking\end{tabular}}       & \textbf{\# Bugs} & \textbf{\begin{tabular}[c]{@{}c@{}}LoC\\ (Avg.)\end{tabular}} & \textbf{\begin{tabular}[c]{@{}c@{}}Test LoC\\  (Avg.)\end{tabular}} & \textbf{\begin{tabular}[c]{@{}c@{}}\# Tests\\ (Avg.)\end{tabular}} & \textbf{\begin{tabular}[c]{@{}c@{}}Coverage\\ (Avg.)\end{tabular}} & \textbf{Crates Involved}      \\ \hline
\textbf{1-50}          & 36                & 25,882                                                       & 13.60             &2137.2  &     79.90\%                   & \parbox{6.5cm}{syn, rand, regex, aho-corasick, num-traits, clap, serde\_json, strsim, time, idna, hashbrown, proc-macro2, smallvec}                         \\ \hline
\textbf{51-100}         & 35                & 25,988                                                       & 13.58             & 409.2 &   86.50\%                                      & \parbox{6.5cm}{percent-encoding, chrono, uuid, textwrap, nom, tokio, hyper, futures, toml}                                          \\ \hline
\textbf{101-150}        & 6                & 26,516                                                       & 23.83           &1578.6  &    90.20\%                                                  & \parbox{6.5cm}{httparse, object, rustc-demangle, rustls, form\_urlencoded, gimli}              \\ \hline
\textbf{151-200} & 3                & 19,173                                                        & 16.67               &216  &   85.50\%                                              & \parbox{6.5cm}{reqwest, num-bignit, rayon}                                         \\ \hline
\textbf{201-250}   & 6                & 3,616                                                        & 9.70                & 27.5  &     80.80\%                                    & \parbox{6.5cm}{bumpalo, filetime, fixedbitset, phf, prost}                                          \\ \hline
\textbf{251-300}         & 3                & 25,572                                                       & 10.33               &78   &   63.80\%                                            & \parbox{6.5cm}{pest, serde-yaml, libm}                  \\ \hline
\textbf{301-350}  & 2                & 44,882                                                       & 10.50                        &17   &  79.10\%                                     & \parbox{6.5cm}{prettyplease, bytemuck}                                 \\ \hline
\textbf{351-400}       & 2                & 1,853                                                        & 12.00             &34.5   & 86.40\%                                   & \parbox{6.5cm}{cargo\_metadata, tinytemplate}                                          \\ \hline
\textbf{401-450}         & 5                & 5,444                                                       & 25.00              &58.7   &   80.70\%                                & \parbox{6.5cm}{tar, plotters, pretty\_assertions, yansi}                                  \\ \hline
\textbf{451-500}   & 4                & 28,760                                                       & 23.25                &6      &   36.50\%                              & \parbox{6.5cm}{crossbeam, brotli-decompressor, indicatif, md5}                                          \\ \midrule
\textbf{Total}        & \textbf{102}      & \textbf{21,805}                                              & \textbf{15.87}                                               & \textbf{456.27} &  \textbf{76.94\%}    & \textbf{51}    \\ \bottomrule  
\end{tabular}}
% \begin{tablenotes}
%   \footnotesize
%   \item[1] The full name of this crate is: aho-corasick.
%   \item[2] The full name of this crate is: percent-encoding.
% \end{tablenotes}
% \end{threeparttable}
% \vspace{-15pt}
\end{table}

\section{Fix Pattern Mining} \label{subsec:pattern_mining}
To comprehensively identify the causes and fix patterns for panic bugs in Rust programs, we analyze and infer data from the Rust implementation code~\cite{rustlang24:online}.
We avoid mining fix patterns solely from \dataset to ensure diversity and comprehensiveness in the repair patterns. 
The root causes of panic errors in \dataset may not be exhaustive; for instance, overflow issues may stem from different operators, such as addition or subtraction, which could be impossible to cover fully through real-world examples. 
In contrast, official implementation code includes all such cases and provides concise examples, making it more suitable for our analysis and collection.

In the Rust compiler source code, panic-related error messages are encapsulated within macros such as \textit{panic\_const!()}.
These macros and their accompanying messages provide the diverse types of panics and their fundamental causes. We compile a summary of these categories and their descriptions in Table~~\ref{tab:root_causes}.
Following this categorization, we examine specific scenarios within each category. Rust compiler developers mark each panic occurrence with the annotation \textit{\# Panics}, which details the causes and circumstances of the panic. 
These annotations are often paired with \textit{\# Safety} or \textit{\# Examples}, providing either fix strategies or bug-triggering examples. 
This annotated information allows us to delineate code-level fix patterns for each panic type.

\begin{table}[htbp]
\centering
\footnotesize
\caption{Root causes of panic bugs derived from Rust compiler source code.}
\label{tab:root_causes}
\vspace{-5pt}
\setlength{\tabcolsep}{3mm} {
\begin{tabular}{l|l}
\toprule
\textbf{Root Causes} & \textbf{Code Examples} \\ \hline
 Unwrap on None/Invalid value &
\parbox{8cm}{
\begin{tabular}{@{}l@{}}
     \texttt{\textcolor{blue}{let} x: Option<i32> = None;} \\
     \texttt{\textcolor{blue}{let} y = x.unwrap();} \texttt{\textcolor{ForestGreen}{// PANIC}} 
\end{tabular}
} \\ \hline
Mixed borrowing &
\parbox{8cm}{
\begin{tabular}{@{}l@{}}
     \texttt{\textcolor{blue}{let} x = Rc::new(RefCell::new(5));} \\
     \texttt{\textcolor{blue}{let} borrow = x.borrow();}
     \texttt{\textcolor{ForestGreen}{// immutable borrow}} \\
     \texttt{\textcolor{blue}{let} borrow\_mut = x.borrow\_mut();} \texttt{\textcolor{ForestGreen}{// PANIC}} 
\end{tabular}
}  \\ \hline

Async functions wrong resume &  
\parbox{8cm}{
\begin{tabular}{@{}l@{}}
    \texttt{\textcolor{blue}{async} \textcolor{blue}{fn} my\_async\_function(cx: \&mut Context<'\_>) \{} \\
    \quad \texttt{\textcolor{blue}{let} mut task = some\_async\_task();} \texttt{\textcolor{ForestGreen}{// starts async task}} \\
    \quad \texttt{\textcolor{blue}{let} result = ready!(task.poll(cx));} \texttt{\textcolor{ForestGreen}{// waits until ready}} \\
    \quad \texttt{ready!(task.poll(cx));} \texttt{\textcolor{ForestGreen}{// PANIC}} \\
    \texttt{\}}
\end{tabular}
} \\ \hline

Arithmetic overflow & 
\parbox{8cm}{
\begin{tabular}{@{}l@{}}
    \texttt{\textcolor{blue}{let} x: i32 = i32::MAX;} \texttt{\textcolor{ForestGreen}{// maximum i32 value}} \\
    \texttt{x + 1;} \texttt{\textcolor{ForestGreen}{// PANIC}}
\end{tabular}
} \\ \hline

Index out of bounds &  
\parbox{8cm}{
\begin{tabular}{@{}l@{}}
    \texttt{\textcolor{blue}{let} array = [1, 2, 3, 4, 5];} \\
    \texttt{array[5];} \texttt{\textcolor{ForestGreen}{// PANIC}}
\end{tabular}
} \\ \hline

Invalid UTF-8 boundary &  
\parbox{8cm}{
\begin{tabular}{@{}l@{}}
    \texttt{\textcolor{blue}{let} s = "™";} \texttt{\textcolor{ForestGreen}{// a UTF\-8 string (3 bytes)}} \\
    \texttt{\&s[1..];} \texttt{\textcolor{ForestGreen}{// PANIC}}
\end{tabular}
} \\ \hline

Division/Modulus by zero & 
\parbox{8cm}{
\begin{tabular}{@{}l@{}}
    \texttt{\textcolor{blue}{let} a = 10; \textcolor{blue}{let} b = 0;} \\
    \texttt{a / b;} \texttt{\textcolor{ForestGreen}{// PANIC}}
\end{tabular}
} \\ \hline

Assertion failed & 
\parbox{8cm}{
\begin{tabular}{@{}l@{}}
    \texttt{assert!(1 == 0);} \texttt{\textcolor{ForestGreen}{// PANIC}}
\end{tabular}
} \\ \hline

Unreachable code &
\parbox{8cm}{
\begin{tabular}{@{}l@{}}
    \texttt{unreachable!();} \texttt{\textcolor{ForestGreen}{// PANIC}}
\end{tabular}
} \\ \hline

Others & 
\parbox{7cm}{
\begin{tabular}{@{}l@{}}
    \texttt{panic!("Here comes a panic!");} \texttt{\textcolor{ForestGreen}{// PANIC}}
\end{tabular}
} \\
\bottomrule
\end{tabular}
}
\vspace{-10pt}
\end{table}

\begin{wrapfigure}{r}{7cm}%靠文字内容的右侧
\centering
% \vspace{-15pt}
  \includegraphics[width=0.95\linewidth]{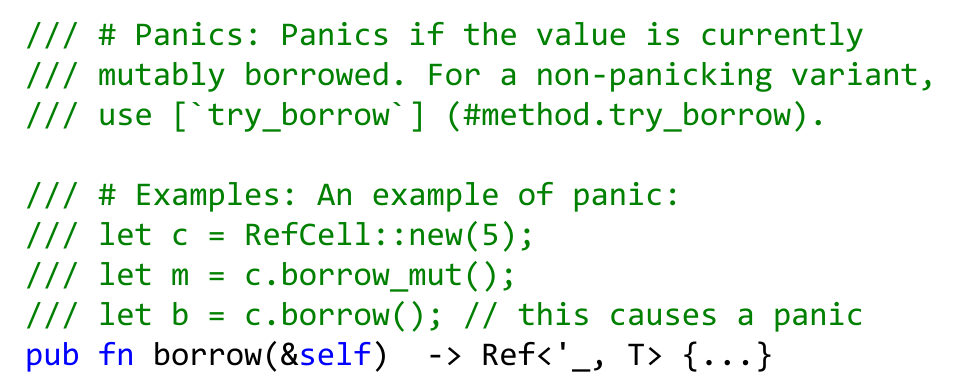}
  \vspace{-8pt}
  \caption{The implementation of \textit{borrow()} function in \textit{RefCell}, including annotation about potential panics.}
  \vspace{-8pt}
  \label{lst:panic}
\end{wrapfigure}
For example, Figure~\ref{lst:panic} showcases a Rust source code snippet annotated to indicate a panic bug related to the ownership mechanism.
% when addressing ownership-related panics in \textit{RefCell}, we examined the \textit{borrow()} method. By reviewing its \textit{\# Panics} annotation, we identified the root cause of the panic. 
The annotation includes an example where the panic is triggered by a double borrowing of a \textit{RefCell} parameter, which is a Rust smart pointer facilitating internal mutability. 
To prevent such panics, the annotation suggests using the \textit{try\_borrow()} method as an alternative approach. 
By systematically reviewing all documented panic instances and their proposed resolutions in the Rust source code, we collect a comprehensive list of fix strategies.

% \begin{listing}[h]
% \begin{minted}[linenos=false,
% fontsize=\scriptsize]{rust}
% /// # Panics: Panics if the value is currently 
% /// mutably borrowed. For a non-panicking variant, 
% /// use [`try_borrow`] (#method.try_borrow).

% /// # Examples: An example of panic:
% /// let c = RefCell::new(5);
% /// let m = c.borrow_mut();
% /// let b = c.borrow(); // this causes a panic
% pub fn borrow(&self)  -> Ref<'_, T> {...}
% \end{minted}
% \vspace{-6pt}
% \caption{The implementation of the \textit{borrow()} function in \textit{RefCell}, including documentation about potential panics.}
% \label{lst:panic}
% \end{listing}

To further delineate fix patterns, we convert code modifications into Abstract Syntax Trees (ASTs). 
By omitting irrelevant tokens, such as variable names, we identify similar AST transformation structures that reveal potential fixes for each identified root cause. 
% In total, we collected 19 fix patterns, of which a representative categorization and corresponding examples are shown in Table~\ref{tab:patterns}, and the complete list of patterns can be found on our website.
In total, we identified 19 fix patterns, which include 34 sub-patterns involving different replacement strategies on one binary operator. A representative categorization along with corresponding examples is presented in Table~\ref{tab:patterns}. The complete list of patterns and sub-patterns is available on our website.

\begin{table}[tp]
\centering
\scriptsize 
\caption{Partial fix patterns derived from the Rust implementation code.}
\vspace{-5pt}
\label{tab:patterns}
\setlength{\tabcolsep}{1mm} {
\begin{tabular}{l|l|l|l}
\toprule
\textbf{Fix Patterns} & \textbf{Code Changes} & \textbf{Interpretation Templates} & \textbf{Example PRs} \\ \midrule
\textbf{Insert Match Unwrapper}  & 
\parbox{4.8cm}{
\begin{tabular}{@{}l@{}}
    \texttt{\textcolor{red}{-\hspace{2mm}x = exp.unwrap(); }} \\
    \texttt{\textcolor{ForestGreen}{+\hspace{2mm}x = match exp.unwrap() \{ }} \\
    \texttt{\textcolor{ForestGreen}{+\hspace{4mm}Some(\_) => \{ exp.unwrap() \} }} \\
    \texttt{\textcolor{ForestGreen}{+\hspace{4mm}\_ => \{ return \} }} \\
    \texttt{\textcolor{ForestGreen}{+\hspace{2mm}\}; }}
    \end{tabular}
} 
% & \parbox{4cm}{Add or revise match arms after unwrapping to handle all possible circumstances, preventing panics caused by inappropriate error handling like unwrap on 'None'.}  
& \parbox{3.7cm}{When unwrapping on [value], add match arms to [variable] after unwrapping to handle all possible circumstances to avoid panics caused by unwrapping on None/Invalid values.}  
& 
\begin{tabular}{@{}l@{}}
    serde\_json \href{https://github.com/serde-rs/json/pull/757}{\textcolor{Violet}{(PR 757)}} \\
    nom \href{https://github.com/rust-bakery/nom/pull/1032}{\textcolor{Violet}{(PR 1032)}}
\end{tabular}
\\ \hline

\textbf{Reorder State Changer} & 
\parbox{4.8cm}{
\begin{tabular}{@{}l@{}}
    \texttt{\textcolor{gray}{//\hspace{1mm}polling operation}} \\
    \texttt{\textcolor{ForestGreen}{+\hspace{1mm}stmt1}}
    \texttt{\textcolor{gray}{//\hspace{1mm}advance statement}} \\
    \texttt{\textcolor{gray}{//\hspace{1mm}other statements}} \\
    \texttt{\textcolor{red}{-\hspace{1mm}stmt1}}
    \texttt{\textcolor{gray}{//\hspace{1mm}state changer}} \\
\end{tabular}
}
& \parbox{3.7cm}{Advance the statement [state changer] to avoid incorrect state resumption after asynchronous functions have completed.}
& \begin{tabular}{@{}l@{}}
    futures \href{https://github.com/rust-lang/futures-rs/pull/2250}{\textcolor{Violet}{(PR 2250)}} \\
    opendal \href{https://github.com/apache/opendal/pull/4013}{\textcolor{Violet}{(PR 4013)}} \\ 
\end{tabular}\\ \hline

\textbf{Delete Second Borrow} & 
\parbox{4.8cm}{
\begin{tabular}{@{}l@{}}
    \texttt{\textcolor{black}{data.borrow()}}
    \texttt{\textcolor{gray}{//\hspace{1mm}immutable borrow}} \\
    \texttt{\textcolor{red}{-\hspace{1mm}data.borrow\_mut()}} 
    % \texttt{\textcolor{gray}{//\hspace{1mm}Mutate-}} \\
    % \texttt{\textcolor{gray}{//\hspace{1mm}-Method Invocation also works}} \\
\end{tabular}
}
& \parbox{3.7cm}{Delete the second mutable borrow of [data] when there exists immutable borrow to avoid ownership violation panics.}
& \begin{tabular}{@{}l@{}}
    StackOverflow \\\href{https://stackoverflow.com/questions/78751978/why-cant-rust-code-process-borrow-checking-at-compile-time/78753157#78753157}{\textcolor{Violet}{(SO 1)}} \\
\end{tabular}\\ \hline

\textbf{Mutate Error Handler} & 
\parbox{4.8cm}{
\begin{tabular}{@{}l@{}}
    \texttt{\textcolor{red}{-\hspace{1mm}x.expect()}}\hspace{2mm}
    \texttt{\textcolor{ForestGreen}{+\hspace{1mm}x.unwrap()}} \\
    \texttt{\textcolor{gray}{//\hspace{1mm}or unwrap\_or\_default/else}}
\end{tabular}
}
& \parbox{3.7cm}{Replace the [original handler] with [new handler] to avoid panics caused by incorrect error handling like [original handler].}
& \begin{tabular}{@{}l@{}}
    clap \href{https://github.com/clap-rs/clap/pull/4480}{\textcolor{Violet}{(PR 4480)}} \\
\end{tabular}\\ \hline

\textbf{Mutate Binary Operator} & 
\parbox{4.8cm}{
\begin{tabular}{@{}l@{}}
    \texttt{\textcolor{red}{-\hspace{2mm}a op b }} \\
    \text{1. Mutate to wrapping/saturating function:} \\
    % \texttt{\textcolor{red}{-\hspace{4mm}a op b }} \\
    \texttt{\textcolor{ForestGreen}{+\hspace{1mm}a.wrapping/saturating\_op(b)}} \\
    \text{2. Mutate to checked function:} \\
    % \texttt{\textcolor{red}{-\hspace{4mm}a op b }} \\
    \texttt{\textcolor{ForestGreen}{+\hspace{2mm}a.checked\_op(b).unwrap()}} \\
    \texttt{\textcolor{gray}{//\hspace{1mm}or unwrap\_or\_default/else}} \\
\end{tabular}
}
% & \parbox{4cm}{Replace basic arithmetic operations with safer operations to handle arithmetic overflow panics.} 
& \parbox{3.7cm}{Replace basic arithmetic operations [operator] with safer operations [call name] to handle arithmetic [operator] overflow panics. Note that [explanation].}
& 
\begin{tabular}{@{}l@{}}
    regex \href{https://github.com/rust-lang/regex/pull/996}{\textcolor{Violet}{(PR 996)}} \\
    chrono \href{https://github.com/chronotope/chrono/pull/1294}{\textcolor{Violet}{(PR 1294)}} \\
    chrono \href{https://github.com/chronotope/chrono/pull/1023}{\textcolor{Violet}{(PR 1023)}} \\
    chrono \href{https://github.com/chronotope/chrono/pull/686}{\textcolor{Violet}{(PR 686)}} \\
\end{tabular}
\\ \hline \hline

\textbf{Insert Range Checker} & 
\parbox{4.8cm}{
\begin{tabular}{@{}l@{}}
    \text{1. Check the index:} \\
    \texttt{\textcolor{ForestGreen}{+\hspace{2mm}if index > arr.len() \{ return \} }} \\
    % \texttt{\textcolor{red}{-\hspace{4mm}x[a]}} \\
    % \texttt{\textcolor{ForestGreen}{+\hspace{4mm}if a > x.len() \{ return \} }} \\
    % \texttt{\textcolor{ForestGreen}{+\hspace{4mm}x[a]}} \\
    \text{2. Check start/end of range if index is range:} \\
    \texttt{\textcolor{ForestGreen}{+\hspace{2mm}if end > x.len() \{ return \} }} or\\
    \texttt{\textcolor{ForestGreen}{+\hspace{2mm}if start >= x.len() \{ return \} }} \\
    % \texttt{\textcolor{red}{-\hspace{4mm}x[a..]}} \\
    % \texttt{\textcolor{ForestGreen}{+\hspace{4mm}if a >= x.len() \{ return \} }} \\
    % \texttt{\textcolor{ForestGreen}{+\hspace{4mm}x[a..]}} \\
    % \text{3. Check the end of range if index is range:} \\
    % \texttt{\textcolor{red}{-\hspace{4mm}x[..b]}} \\
    % \texttt{\textcolor{ForestGreen}{+\hspace{4mm}if b > x.len() \{ return \} }} \\
    % \texttt{\textcolor{ForestGreen}{+\hspace{4mm}x[..b]}}
    \text{3. Check whether start $>$ end if index is range:} \\
    \texttt{\textcolor{ForestGreen}{+\hspace{2mm}if start > end \{ return \}}} \\
\end{tabular}
}
% & \parbox{4cm}{Implement range checking for the range index of indices before their usage, preventing panics caused by out-of-bounds errors.} 
& \parbox{3.7cm}{Implement range checking for the [index] of indices [array name] to determine whether [condition], avoiding index out of bounds or exceed the boundary.} 
& \begin{tabular}{@{}l@{}}
    idna \href{https://github.com/servo/rust-url/pull/658}{\textcolor{Violet}{(PR 658)}} \\
    idna \href{https://github.com/servo/rust-url/pull/655}{\textcolor{Violet}{(PR 655)}} \\
\end{tabular}
\\ \hline

\textbf{Mutate Index Expression} & 
\parbox{4.8cm}{
\begin{tabular}{@{}l@{}}
     \texttt{\textcolor{red}{-\hspace{1mm}array[index1]}}\hspace{2mm}
     \texttt{\textcolor{ForestGreen}{+\hspace{1mm}array[index2]}}
\end{tabular}
} 
% & \parbox{4cm}{Mutate the index expression to avoid panics caused by out-of-bounds errors resulting from incorrect indices.} 
& \parbox{3.7cm}{Mutate [index] in indices [array name], avoiding index out of bounds or exceeding the boundary.} 
& \begin{tabular}{@{}l@{}}
    textwrap \href{https://github.com/mgeisler/textwrap/pull/391}{\textcolor{Violet}{(PR 391)}} \\
\end{tabular}\\ \hline \hline

\textbf{Mutate Condition} & 
\parbox{4.8cm}{
\begin{tabular}{@{}l@{}}
    % \text{1. Add additional condition:} \\
    \texttt{\textcolor{red}{-\hspace{1mm}if cond1 }}\hspace{2mm}
    \texttt{\textcolor{ForestGreen}{+\hspace{1mm}if cond2 \&\& cond1 }} \\
    \texttt{\textcolor{gray}{//\hspace{1mm}add after if-let expression}}
    % \text{2. Add if expression after if-let expression:} \\
    % \texttt{\textcolor{red}{-\hspace{1mm}if let x = pat}}\hspace{2mm}
    % \texttt{\textcolor{ForestGreen}{+\hspace{1mm}if let x = pat if cond }}
\end{tabular}
} 
% & \parbox{4cm}{Adjust conditions within if statements to better handle edge cases or boundary conditions, thus preventing panics due to unexpected values or states.} 
& \parbox{3.7cm}{Adjust conditions within if statements to check whether [condition].} 
& \begin{tabular}{@{}l@{}}
    idna \href{https://github.com/servo/rust-url/pull/865}{\textcolor{Violet}{(PR 865)}} \\
\end{tabular}
\\ \hline

\textbf{Insert Unsafe Block} & 
\parbox{4.8cm}{
\begin{tabular}{@{}l@{}}
    \texttt{\textcolor{gray}{//\hspace{1mm}necessary condition}} \\
    \texttt{\textcolor{red}{-\hspace{1mm}exp1}}\hspace{2mm}
    \texttt{\textcolor{ForestGreen}{+\hspace{1mm}unsafe \{ exp2 \}}}
\end{tabular}
}
% & \parbox{4cm}{Insert an unsafe block to circumvent Rust's safety checks when certain preconditions are met, such as bypassing UTF-8 checks when boundary check has existed.} 
& \parbox{3.7cm}{Insert an unsafe block when [precondition] is met to change the behaviour of [variable].} 
& 
\begin{tabular}{@{}l@{}}
    nom \href{https://github.com/rust-bakery/nom/pull/370}{\textcolor{Violet}{(PR 370)}} \\
\end{tabular}\\ \hline \hline

\begin{tabular}{@{}l@{}}\textbf{Mutate Method} \\ \textbf{Invocation} \end{tabular} & 
\parbox{4.8cm}{
\begin{tabular}{@{}l@{}}
     \texttt{\textcolor{red}{-\hspace{1mm}x.y([params])}} \hspace{2mm}
     \texttt{\textcolor{ForestGreen}{+\hspace{1mm}x.z([params])}}
\end{tabular}
} 
& \parbox{3.7cm}{Replace the original call [call name] with another [new call name] with the same parameters.} 
&
\begin{tabular}{@{}l@{}}
    hyper \href{https://github.com/hyperium/hyper/pull/2410}{\textcolor{Violet}{(PR 2410)}} \\
     serde\_json \href{https://github.com/serde-rs/json/pull/493}{\textcolor{Violet}{(PR 493)}} \\
\end{tabular}
\\ \hline 

\textbf{Insert Call Invocation} & 
\parbox{4.8cm}{
\begin{tabular}{@{}l@{}}
     \texttt{\textcolor{red}{-\hspace{1mm}x.y()}}\hspace{2mm}
     \texttt{\textcolor{ForestGreen}{+\hspace{1mm}x.y().z()}}
\end{tabular}
} 
& \parbox{3.7cm}{Add new method call [call name] to [variable].} 
& 
\begin{tabular}{@{}l@{}}
    chrono \href{https://github.com/chronotope/chrono/pull/1254}{\textcolor{Violet}{(PR 1254)}} \\
    % chrono \href{https://github.com/chronotope/chrono/pull/1024/files}{\textcolor{Violet}{(PR 1024)}} \\
    nom \href{https://github.com/rust-bakery/nom/pull/1618}{\textcolor{Violet}{(PR 1618)}} \\
\end{tabular}
\\ 

\bottomrule

\end{tabular}
}
\vspace{-5pt}
\end{table}

\section{Automated Panic Bugs Fixing Tool}\label{subsec:tool}

Based on the mined fix patterns, we design a pattern-based automated panic bugs fixing tool, namely \tool.
\tool first locates suspicious expressions based on the stacktrace information output by the compiler and the dependency flow analysis of the original program. 
Then, combined with the semantic information of the errors, \tool generates a series of patches for each fault location. 
Finally, \tool sorts all the patches with verification and matching scores calculation, and outputs the top-5 ranked patches along with the corresponding natural language interpretation.

\subsection{Fault Localization}

When a panic occurs, Rust initiates the unwinding process, meticulously tracing back up the stack to ensure the cleanup process.
% Concurrently, an error message, rich with crucial details, is generated. 
The error message is further supplemented with stacktrace information, and based on it, \tool initially extracts suspicious locations, including file paths, column and row numbers.
Considering that the precise locations of bugs might not be explicitly disclosed in the stacktrace~\cite{schroter2010stacktrace}, prior research suggests that the actual buggy location is apt to exhibit structural resemblances to the expressions detailed in the stacktrace. This indicates that the exact location could be involved in other expressions and is dependent on those specified within the stacktrace.

To capture the dependency relationship, we construct a Code Element Dependency Graph (CEDG) by utilizing the High-level Intermediate Representation (HIR), which serves as one of the intermediary forms during Rust's compilation process.
% Specifically, we undertake a traversal of the assign statements or expressions within the HIR. 
% \textcolor{red}{For \texttt{let} statements, the pattern and corresponding type are contingent on the initializer expression, implying that data flows from the initializer expression to the pattern. }
For the assignment statements or expressions, the variables on the left-hand side values are dependent on the right-hand side values. 
When it comes to function invocations, there may exist dependencies in the parameters of functions, so we delve deeper into the function to uncover additional dependencies. 
Leveraging the transitivity of the dependency relationship, we iteratively construct the CEDG.
Then, based on the constructed dependency relationship, we first define the confidence score of each localized element $e_i$ as follows:

\begin{equation}\label{eq:simi_loc}
\small
    Con_i = 1 - \frac{\min(Dist(\text{stacktrace}, e_i), \lambda)}{\lambda}, \quad \lambda \geq 1 \\
\end{equation}

where $Dist(\text{stacktrace}, e_i)$ denotes the shortest distance on the dependency graph between element $e_i$ and any of the elements in the stacktrace. 
A constant $\lambda$ is used to normalize the confidence score thus ensuring it is scaled within a meaningful range.
In our work, we set $\lambda=2$ according to an existing study~\cite{laura2014dep}, which has proved that it has the optimal performance.
It involves incorporating both the precise locations and the code elements dependent on those identified, ensuring a more comprehensive analysis.

In addition to the confidence score, the presence of suspicious files within the stacktrace can also impact fault localization. 
On the one hand, files that frequently appear in the stacktrace tend to be more critical to the execution path leading to the error, making them more likely to be associated with the fault. 
On the other hand, within an expanded stacktrace, a shallower depth suggests a stronger connection to the error's origin~\cite{wong2014stacktrace}. 
Thus, \tool computes the suspicion score $Loc_i$ for each location as follows:

\begin{equation}\label{eq:score_loc}
\small
Loc_i = N \times \left( 1/D + Con_i \right)
\end{equation}

where $N$ denotes the number of times a particular file of $e_i$ occurs within the stacktrace, and $D$ represents the depth of the location in the stacktrace sequence.
For each buggy source code and the corresponding stacktrace, \tool calculates the score for each element $e_i$ after constructing a dependency graph to determine the ranking of a suspect location.

% It appears that despite numerous approaches claiming performance improvements, there has been little adoption of Automatic Program Repair research or practices. This lack of adoption could be attributed to issues with coarse-grained fault localization, as discussed by Koyuncu et al.~\cite{koyuncu2019ifixr}. It's crucial to emphasize that, in contrast, \tool employs fault localization at the expression level. This detail will be further elaborated in Section~\ref{subsec:rq1}. Such a granular approach to fault localization significantly enhances the applicability of our method in real-world scenarios.

\subsection{Pattern-based Patch Generation} \label{subsubsec:generation}

Based on the suspicious locations, \tool iteratively attempts to match each location with the mined fix patterns, which are illustrated in Section~\ref{subsec:pattern_mining}. 
% Specifically, \tool parses the suspicious file to AST, and perform transformations according to the localized expression with row and column number.
Considering that there may be multiple suspicious locations and matched patterns, resulting in multiple combinations of fixes, we sort them based on validation.

\textbf{Patch validation.} 
% The comprehensive process, encompassing both fault localization and patch generation, yields multiple patch candidates for each identified Rust panic bug. 
% The generation of high-quality patches plays a crucial role in the overall quality of repairs, highlighting the importance of not only identifying faults but also rectifying them effectively. 
% Moreover, addressing the issue of patch overfitting is essential to ensure the selection of genuinely effective patches, avoiding solutions that merely succeed in passing the current test suite without genuinely correcting the underlying bug.
For the APR task, it is important to guarantee that the generated patches repair the bug as well as do not affect the original semantics.
In our work, we incorporate \textit{cargo-test}, a built-in tool in Rust's package manager, to automate the execution of regression tests. 
After performing the validation, the patches would have several cases: 
(1) The panic is eliminated and all the test cases execute with the same result as before.
Our goal is to generate these patches, which indicate the most likely correct fix.
(2) The panic is eliminated but the execution results of some test cases are not consistent with the original program.
This suggests that the patch may have introduced logical modifications that cause the semantic inconsistency.
(3) The panic is not eliminated and we don't evaluate the regression testing because it's not a correct fix.
% The validation results are further applied to \tool to output the final patch ordering.
By regression testing, we can effectively filter out patches that might introduce new errors or exhibit overfitting to test suites, thereby enhancing the reliability and robustness of fixing.

%%%%%%%%%%%%%%%%%%%%%%%%%%%%%%%%%%%%%%%%%%%%

\textbf{Patch prioritization.}
For each suspicious location, \tool iteratively employs the mined fix patterns.
If the corresponding AST has a structure that matches a specific pattern, it is applied to generate a patch; otherwise, it is discarded.
Note that even after a pattern has been applied to an expression, additional patterns might be identified as one delves deeper into the AST. Consequently, a suspicious location may correspond to more than one resulting patch.
For all the generated patches, \tool combines the scores and validation results mentioned above to perform the patch ordering.
% Specifically, three factors are used to guide the prioritization of patches: the confidence score used for error localization, the similarity score of the patch to the error cue, and the validation results of the regression test.
% For all the generated patches, \tool aggregates the scores and validation results mentioned above to order the patches. 
Specifically, patch prioritization is guided by three factors: (1) the confidence score for fault localization $Con_i$, (2) the similarity score of the patch's interpretation, the templates of which are shown in Table~\ref{tab:patterns}, to the error messages, and (3) the results from the regression test validation.
% The similarity of a patch is the degree of similarity between the natural language interpretation produced when applying each patch, the templates of which are shown in Table~\ref{tab:patterns}, and the actual panic error alert. 
% A higher similarity score suggests that the pattern is more likely to meet the actual requirements.
\tool ranks each patch based on the cumulative sum of the two scores. 
For patches that achieve identical scores, preference is given to those that successfully pass regression testing.

%\textbf{Output.}
Finally, \tool outputs a top-5 ranked list of mixed-form fixing recommendations. 
To address patch interpretability issues~\cite{huang2023survey}, \tool provides each code patch alongside its natural language explanation, as detailed in Table~\ref{tab:patterns}. 
This description ensures developers receive a clear understanding of the modifications and the patterns applied in each patch.

% \begin{figure}[htbp]
%     \centering
%     \includegraphics[width=0.9\linewidth]{fig/fig_example.pdf}
%     \vspace{-10pt}
%     \caption{A panic bug-triggering example and its corresponding patches.}
%     \label{fig:example}
%     \vspace{-5pt}
% \end{figure}

\begin{figure}[htbp]
    \centering
    \includegraphics[width=\linewidth]{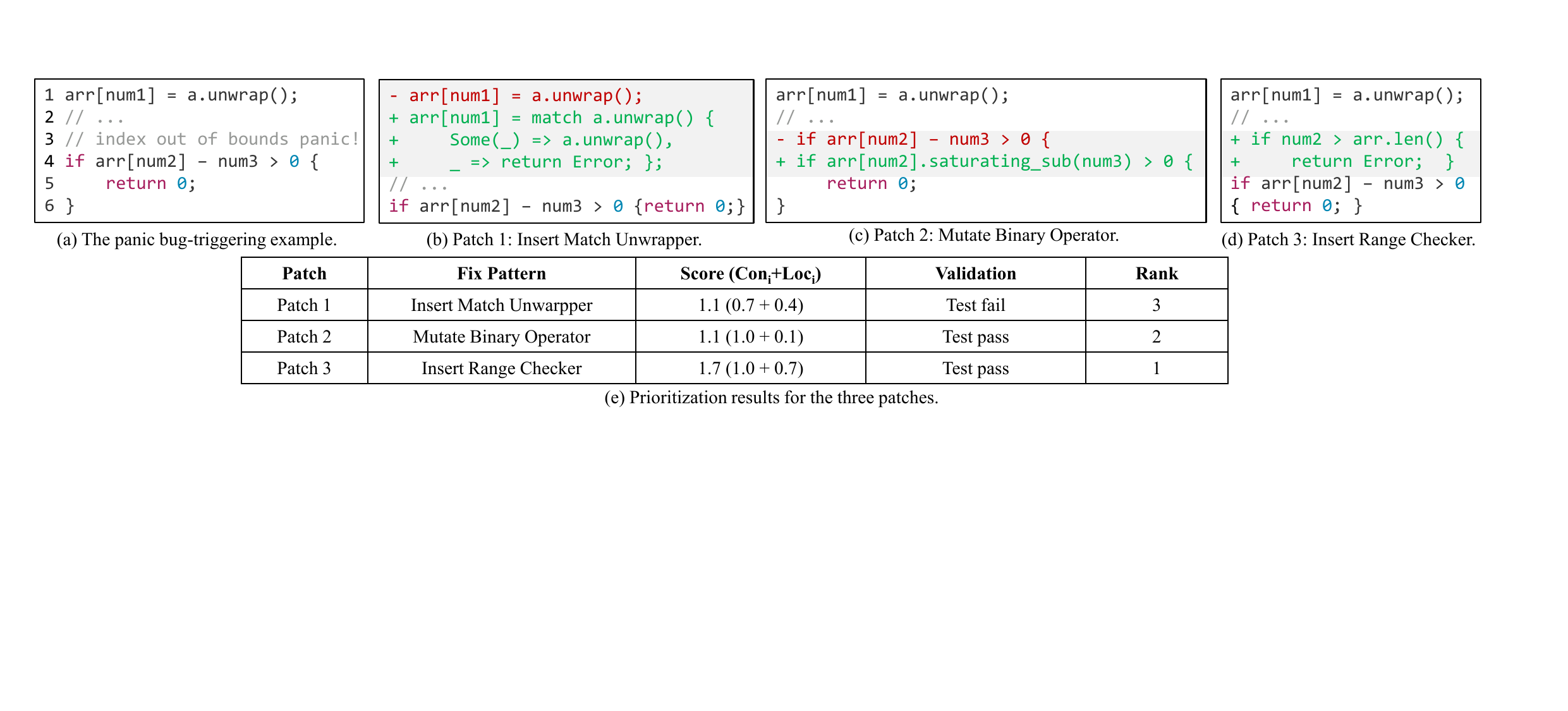}
    \vspace{-10pt}
    \caption{A panic bug-triggering example and its corresponding patches.}
    \label{fig:example}
    \vspace{-5pt}
\end{figure}

\textbf{A running example.}
Figure~\ref{fig:example} (a) presents a code example that triggers a panic bug caused by an index out-of-bounds error. 
Assuming that \tool has identified lines $1$ and $4$ as suspicious locations, with confidence scores of $0.7$ and $1$, respectively, as determined by Equation~\ref{eq:score_loc}.
Subsequently, \tool iteratively applies each pattern to these locations. 
For line $1$, \tool identifies a method call of \texttt{unwrap()}, thus the \textit{Insert Match Unwrapper} pattern is applied, and \textit{patch 1} is depicted in Figure~\ref{fig:example} (b).
For line $4$, \tool initially identifies a binary expression containing a subtraction operator, and the \textit{Mutate Binary Operator} pattern is employed, as illustrated as \textit{patch 2} in Figure~\ref{fig:example} (c). 
Further, \tool analyzes the deeper structure and identifies an index expression, for which the \textit{Insert Range Checker} pattern is applied to \texttt{arr[num2]}. 
\textit{Patch 3} is depicted in Figure~\ref{fig:example} (d).
% \begin{table}[htbp]
% \centering
% \footnotesize
% \caption{Prioritization results for the three patches.}
% \label{tab:example}
% \vspace{-5pt}
% \setlength{\tabcolsep}{0.5mm} {
% \begin{tabular}{|c|c|c|c|c|}
% \hline
% \textbf{Patch}  &  \textbf{Fix Pattern} & \textbf{Score ($Con_i$+$Loc_i$)} & \textbf{Validation} & \textbf{Rank} \\ \hline
% Patch 1  & Insert Match Unwrapper & 1.1 (0.7 + 0.4)     & Test fail  & 3  \\ \hline
% Patch 2  & Mutate Binary Operator & 1.1 (1.0 + 0.1)    & Test pass   & 2   \\ \hline
% Patch 3  & Insert Range Checker   & 1.7 (1.0 + 0.7)     & Test pass   & 1   \\ \hline
% \end{tabular}}
% \vspace{-5pt}
% \end{table}
The prioritization of patches is shown in Figure~\ref{fig:example} (e). \textit{Patch 3} ranks first with the highest score of $1.7$, having passed the regression tests. 
\textit{Patch 2} is second because, although it scored the same as \textit{patch 1}, it passed the regression tests while \textit{patch 1} failed. 
Consequently, \tool outputs this ranked patch list with their interpretations, organized by scores and test results.

\subsection{Tool Evaluation}
To the best of our knowledge, there currently exists no dedicated tool for the automated repair of Rust panic bugs. 
Therefore, we conducted an evaluation on the \dataset dataset to validate the effectiveness of \tool. We mainly compare the effectiveness of \tool with LLM-based approaches.
We follow a standardized process when applying ChatGPT 4.0 as a baseline method for fault localization and program repair.
For each target, we upload a zip archive containing the entire Rust project, the test case triggering the panic, and relevant panic information. 
We then use a consistent prompt template, as shown below:

\begin{framed}
\footnotesize \textsf{I have uploaded a source code package of a Rust crate \textbf{[crate-name]}, \textbf{[crate-zip]}. When using this crate with the following test case: \textbf{[main.rs]}, it went panic, the panic information is as follows: \textbf{[panic\_info]}.}

\footnotesize \textsf{\textbf{[Location Prompt]} + Based on the locations, provide the top 5 \textbf{fixing code} to fix this panic. Don't explain anything about code, just show me the suspicious locations and patches.}
\end{framed}

As for the \textbf{[Location Prompt]}, because we verify the correctness of the generated patches with/without the correct error location, the template are specifically as follows:

\begin{framed}

\begin{itemize}[leftmargin=*]
    \item \footnotesize \textbf{Without Perfect Location}: \textsf{Please give me the top 5 \textbf{suspicious fault locations} in the crate, including their path, line number and column number.} 
    \item \footnotesize \textbf{With Perfect Location}: \textsf{The error location is: \textbf{[perfect-location]}.}
\end{itemize}

\end{framed}

Because ChatGPT may fail to produce bug localization or repair patches as instructed, we employ a baseline involving multiple interactions with ChatGPT. 
Initially, we assess ChatGPT's output for completeness and iterate the query process up to three times, correcting for any missing information. 
If, after three attempts, localization or repair details remain unattained, we classify the effort as a ChatGPT repair failure. 
The inquiry prompts for missing information are as follows:

\begin{framed}

\begin{itemize}[leftmargin=*]
\item \footnotesize \textsf{\textbf{Localization}: Please provide me directly with the exact \textbf{file path}, \textbf{code line} and \textbf{column numbers} where the error was located.}

\item \footnotesize \textsf{\textbf{Fixing}: Based on the mislocalization information you provided, please provide me with the specific code to fix it.}
\end{itemize}

\end{framed}

\subsubsection{Effectiveness of Fault Localization} \label{subsec:rq1}
For the task of bug localization, while existing techniques typically rely on additional natural language descriptions from bug reports or test suites~\cite{yoo2012regression}, our tool, \tool, solely utilizes the error message from the Rust compiler and the accompanying stacktrace information. 
The absence of extra data makes some existing techniques unsuitable as baselines for our approach. 
Consequently, we select three categories of general methods as baseline techniques to compare with the fault localization capabilities of \tool.
% Consequently, we devised three baseline methods based on conventional localization approaches:
% (1) \textit{panic}: Directly using the location indicated in the panic error message.
% (2) \textit{random}: Randomly selecting a suspicious location from the stacktrace.
% (3) \textit{similarity}: Calculating textual similarity between the error message and each code location; the highest similarity score indicates the fault location.

\textbf{(1) Conventional Localization Approaches.} When only error messages and stacktraces are available, fault localization can be deduced from these elements\cite{jiang2012stack, laura2014dep, wong2014stacktrace}.
% We outline three fundamental techniques below: \liu{add cites}
\begin{itemize}
\item \textit{Panic}: When a panic occurs, the compiler outputs an error message indicating the specific line of code where the exception occurred. This line is then considered as the fault location.
\item \textit{Random}: The compiler outputs stacktrace information that includes a list of suspicious code locations. From this list, a location is randomly selected as the detected fault location.
\item \textit{Similarity}: We evaluate the textual error message and identify the fault location by finding the code line with the highest similarity score.
\end{itemize}

\textbf{(2) Spectrum-based Localization Approaches.} 
Spectrum-based fault localization (SBFL)~\cite{abreu2007spectrum} approaches analyze the execution traces of a program by examining which components of the program were executed when a failure occurred and which were not. 
% The spectrum refers to the range of execution traces collected during test runs. 
By applying statistical analysis to these execution spectra, SBFL effectively pinpoints the components that are most likely to be responsible for software failures. 
In our experiments, we collect test cases for each crate in \dataset, the majority of which passed; those that triggered a panic bug were classified as failed cases.
Based on the calculated scores from the spectral analysis, we sort all the lines of code and then select the top 5 locations as the fault localization results.
% Due to slight variations in how the spectrum is calculated, we select the two most commonly used metrics as our baselines.
\begin{itemize}
\item \textit{Tarantula}: The Tarantula method~\cite{jones2002tarantula} calculates the suspicious location by comparing the ratios of passed to failed test cases that execute a given code line. A higher ratio of failed to total test executions in a component increases its likelihood of being faulty. We rank the ratio of each code line as the most likely fault candidates.
\item \textit{Ochiai}: The Ochiai algorithm~\cite{meyer2004comparison} applies the cosine similarity measure between failed test cases and the execution of given code lines. The score is derived from the intersection of failing test cases with the components they execute, normalized by the square root of the product of the total number of failures and the number of times a component is executed during those failures. 
\end{itemize}

\textbf{(3) LLM-based Localization Approaches.} It is feasible to directly submit source code, error messages, and requirements to an LLM. 
% In response, the LLM could generate corresponding fault localization results.
\begin{itemize}
\item \textit{ChatGPT-4 (GPT-4)}: Due to ChatGPT-3.5's limitations in handling files, we apply the most popular commercial software ChatGPT 4.0 as a baseline.
\item \textit{Multiple rounds of inquiries with ChatGPT-4 (GPT4-multi)}: When interacting with ChatGPT 4.0, we inquire multiple rounds with additional questions manually.
\end{itemize}

To comprehensively assess the effectiveness of fault localization through various techniques, we employ three granularity levels for locating panics: file, statement, and expression levels. 
As illustrated in Table~\ref{tab:rq1}, the localization accuracy of \tool at all three granularities is higher than the other baselines, which indicates the effectiveness of \tool.
In addition, all methods perform better on the small-scale dataset compared with the large-scale dataset. 
This result is as expected, because on large-scale projects, the stack unfolding information could be more complex and harder to analyze the dependency for fault localization.
The effectiveness of the random method is relatively close to \tool, especially on small-scale datasets and top-5 prediction results, which suggests that if the localization is in the stack-expanded list, random selection can be effective, but as soon as in-depth dependency analyses are required on large-scale dataset, the random method fails outright.
The similarity-based selection techniques are the least effective in terms of fine-grained localization accuracy, implying that relying on natural language alone does not accurately capture the semantic information of complex practical code, resulting in low localization accuracy.

As for comparing with LLM-based approaches, the localization accuracy of ChatGPT4 is basically the same for single and multi-round conversations due to a limited understanding of an entire Rust project. This limitation is evident in the significant drop in success rate as the size of the Rust project increases. 
Throughout our experiments, we observed instances where ChatGPT failed to parse files accurately, providing only general suggestions that are impractical for effective fault localization. These observations highlight the inconsistency and limitations of relying on LLMs for accurate bug localization in complex codebases.

For spectrum-based methods, both Ochiai and Tarantula methods exhibited low accuracy, particularly within larger datasets, where Ochiai, for instance, showed zero accuracy in the top-1 ranking across all levels. 
This may because that panic bugs in Rust program constitute a unique category of errors, usually not originating from traditional issues in loops or conditional logic. The unit test cases within each crate aim to validate logical functionality, thus testing pass and fail conditions that may not align with detecting panic-related bugs. This mismatch could reduce the effectiveness of spectrum-based techniques. 
In addition, facing with real-world and large-scale programs that contain thousands of lines code, applying spectrum-based techniques incurs considerable time overheads, further limiting their practicality. 
% In fact, the time consumption of these spectrum-based methods is significantly higher compared to our approach. 
For example, calculating the spectrum score for a project could takes more than 3 hours in average for the large-scale dataset, while \tool takes about 1 minute in average. 

\begin{table}[ht]
\centering
\caption{The correctness of tools on \dataset datasets, with different granularities of error localization.}
\vspace{-5pt}
\label{tab:rq1}
\scriptsize
\setlength{\tabcolsep}{1mm}{
\begin{tabular}{cc|ccc|ccc||ccc}
\toprule
\multicolumn{2}{c|}{\multirow{2}{*}{\textbf{Tool}}} & \multicolumn{3}{c|}{\textbf{\dataset-Small (61)}}     & \multicolumn{3}{c||}{\textbf{\dataset-Large (41)}}   & \multicolumn{3}{c}{\textbf{Total (102)}}    \\ %\cline{3-11} 
&& \textbf{File}  & \textbf{Stmt}  & \textbf{Expr}  & \textbf{File}  & \textbf{Stmt}  & \textbf{Expr}  & \textbf{File}  & \textbf{Stmt} & \textbf{Expr}  \\ \midrule
\multirow{8}{*} {Top-1} & panic        & 41 (67.2\%) & 30 (49.2\%) & 30 (49.2\%) & 22 (53.7\%) & 10 (24.4\%) & 10 (24.4\%) & 63 (61.8\%) & 40 (39.2\%) & 40 (39.2\%) \\ 
& random       & 33 (54.1\%) & 14 (23.0\%)  & 13 (21.3\%)  & 14 (34.1\%) & 7 (17.1\%)  & 6 (14.6\%) & 47 (46.1\%) & 21 (20.6\%) & 19 (18.6\%) \\ 
& similarity          & 13 (21.3\%)  & 0 (0\%)     & 0 (0\%)     & 7 (17.1\%)  & 1 (2.4\%)   & 0 (0\%)   & 20 (19.6\%) & 1 (1.0\%)   & 0 (0\%)   \\ \cline{2-11}
& Ochiai  & 0 (0\%) & 0 (0\%)  & 0 (0\%)  & 0 (0\%) & 0 (0\%)  & 0 (0\%) & 0 (0\%) & 0 (0\%) & 0 (0\%) \\ 
& Tarantula  & 1 (1.6\%) & 1 (1.6\%)  & 1 (1.6\%)  & 0 (0\%) & 0 (0\%)  & 0 (0\%) & 1 (1.0\%) & 1 (1.0\%) & 1 (1.0\%) \\ \cline{2-11}
& GPT-4        & 42 (68.9\%) & 28 (45.9\%) & 19 (31.1\%) & 10 (24.4\%)  & 4 (9.8\%)  & 3 (7.3\%) & 52 (51.0\%) & 32 (31.4\%) & 22 (21.6\%) \\ 
& GPT4-multi   & 43 (70.5\%) & 28 (45.9\%) & 19 (31.1\%) & 10 (24.4\%)  & 4 (9.8\%)  & 3 (7.3\%) & 53 (52.0\%) & 32 (31.4\%) & 22 (21.6\%) \\ 
&\textbf{\tool}  & \textbf{45 (73.8\%)} & \textbf{33 (54.1\%)} & \textbf{32 (52.5\%)} & \textbf{22 (53.7\%)} & \textbf{13 (31.7\%)} & \textbf{13 (31.7\%)} & \textbf{67 (65.7\%)} & \textbf{46 (45.1\%)} & \textbf{45 (44.1\%)} \\ \midrule
\multirow{7}{*} {Top-3} & random       & 46 (75.4\%) & 29 (47.5\%) & 29 (47.5\%) & 17 (41.5\%) & 8 (19.5\%)  & 8 (19.5\%) & 63 (61.8\%) & 37 (36.3\%) & 37 (36.3\%) \\ 
& similarity          & 21 (34.4\%)  & 1 (1.6\%)     & 1 (1.6\%)     & 10 (24.4\%)  & 1 (2.4\%)   & 0 (0\%)   & 31 (30.4\%) & 2 (2.0\%) & 1 (1.0\%)   \\ \cline{2-11}
& Ochiai  & 3 (5.0\%) & 3 (5.0\%)  & 3 (5.0\%)  & 0 (0\%) & 0 (0\%)  & 0 (0\%) & 3 (2.9\%) & 3 (2.9\%) & 3 (2.9\%) \\ 
& Tarantula  & 2 (3.3\%) & 2 (3.3\%)  & 2 (3.3\%)  & 0 (0\%) & 0 (0\%)  & 0 (0\%) & 2 (2.0\%) & 2 (2.0\%) & 2 (2.0\%) \\ \cline{2-11}
& GPT-4        & 43 (70.5\%) & 29 (47.5\%) & 20 (32.8\%) & 12 (29.3\%)  & 4 (9.8\%)  & 3 (7.3\%) & 55 (53.9\%) & 33 (32.4\%) & 23 (22.5\%) \\ 
& GPT4-multi   & 44 (72.1\%) & 29 (47.5\%) & 20 (32.8\%) & 12 (29.3\%)  & 4 (9.8\%)  & 3 (7.3\%) & 56 (54.9\%) & 33 (32.4\%) & 23 (22.5\%) \\ 
&\textbf{\tool}  & \textbf{49 (80.3\%)} & \textbf{38 (62.3\%)} & \textbf{37 (60.7\%)} & \textbf{28 (68.3\%)} & \textbf{19 (46.3\%)} & \textbf{18 (43.9\%)} & \textbf{77 (75.5\%)} & \textbf{57 (55.9\%)} & \textbf{55 (53.9\%)} \\ \midrule
\multirow{7}{*} {Top-5} & random       & 48 (78.7\%) & 33 (54.1\%) & 33 (54.1\%) & 22 (53.7\%) & 13 (31.7\%) & 12 (29.3\%) & 70 (68.6\%) & 46 (45.1\%) & 45 (44.1\%) \\ 
& similarity          & 25 (41.0\%)  & 1 (1.6\%)     & 1 (1.6\%)     & 13 (31.7\%) & 1 (2.4\%)   & 0 (0\%)  & 38 (37.3\%) & 2 (2.0\%) & 1 (1.0\%)   \\\cline{2-11}
& Ochiai  & 9 (14.8\%) & 9 (14.8\%)  & 9 (14.8\%)  & 0 (0\%) & 0 (0\%)  & 0 (0\%) & 9 (8.8\%) & 9 (8.8\%) & 9 (8.8\%) \\ 
& Tarantula  & 6 (9.8\%) & 6 (9.8\%)  & 6 (9.8\%)  & 1 (2.4\%) & 1 (2.4\%)  & 1 (2.4\%) & 7 (6.9\%) & 7 (6.9\%) & 7 (6.9\%) \\ \cline{2-11}
& GPT-4        & 46 (75.4\%) & 30 (49.2\%) & 21 (34.4\%) & 14 (34.1\%) & 4 (9.8\%)  & 3 (7.3\%) & 60 (58.8\%) & 34 (33.3\%) & 24 (23.5\%) \\
& GPT4-multi   & 47 (77.0\%) & 30 (49.2\%) & 21 (34.4\%) & 14 (34.1\%) & 4 (9.8\%)  & 3 (7.3\%) & 61 (59.8\%) & 34 (33.3\%) & 24 (23.5\%) \\ 
&\textbf{\tool}  & \textbf{49 (80.3\%)} & \textbf{38 (62.3\%)} & \textbf{37 (60.7\%)} & \textbf{29 (70.7\%)} & \textbf{20 (48.8\%)} & \textbf{19 (46.3\%)} & \textbf{78 (76.5\%)} & \textbf{58 (56.9\%)} & \textbf{56 (54.9\%)} \\ \bottomrule
\end{tabular}}
%\vspace{-10pt}
\end{table}

\subsubsection{Fixing Effectiveness}
Referring to existing research on APR~\cite{qi2015plausible}, we categorize the generated patches into three types and verify their proportion: (1) panic-eliminated patches, where the panic bug is resolved but some regression test cases fail; (2) plausible patches, where the panic bug is successfully fixed and all test cases pass; and (3) correct patches, which are plausible patches that have also been manually verified for semantic correctness.
The evaluation results are shown in Table~\ref{tab:rq2.1}.
Overall, we can conclude that \tool significantly surpasses the performance of automated repairs conducted by GPT-4. 
Notably, within the large-scale dataset, GPT-4 fails to generate any viable patches, the reason of which is closely tied to GPT-4's challenges with fault localization, as elaborated in Section~\ref{subsec:rq1}. 
Moreover, we find that LLM-based methods frequently overlooks the code context when generating patches, resulting in the generated program failing to compile instead.
For example, it may attempt to resolve panics by altering a method's return type directly or by incorporating new logic, which results in both compilation and semantic errors.

\begin{table}[htbp]
\centering
\footnotesize
\caption{The number of panic-eliminated/plausible/correct patches of different tools on \dataset.} 
\vspace{-5pt}
\label{tab:rq2.1}
\setlength{\tabcolsep}{0.6mm}{ 
\begin{tabular}{cc|cc|cc|cc|cc|cc|cc||cc|cc|cc}
\toprule
\multicolumn{2}{c|}{\multirow{2}{*}{\textbf{Tool}}} & \multicolumn{6}{c|}{\textbf{\dataset-Small (61)}}                                         & \multicolumn{6}{c||}{\textbf{\dataset-Large (41)}}          & \multicolumn{6}{c}{\textbf{Total (102)}}                \\ %\cline{3-8} 
&   & \multicolumn{2}{c|}{\textbf{Eliminated}} & \multicolumn{2}{c|}{\textbf{Plausible}}      & \multicolumn{2}{c|}{\textbf{Correct}}       & \multicolumn{2}{c|}{\textbf{Eliminated}} & \multicolumn{2}{c|}{\textbf{Plausible}}   & \multicolumn{2}{c||}{\textbf{Correct}}  & \multicolumn{2}{c|}{\textbf{Eliminated}} & \multicolumn{2}{c|}{\textbf{Plausible}}  & \multicolumn{2}{c}{\textbf{Correct}} \\ \midrule
%\multirow{3}{*} {Top-1}
\multirow{3}{*}{\begin{sideways}Top-1\end{sideways}} 
& GPT-4          
& 6 & \cellcolor[rgb]{ .930,  .930,  .930}9.8\%  & 6 & \cellcolor[rgb]{ .930,  .930,  .930}9.8\% & 0  & \cellcolor[rgb]{ .930,  .930,  .930}0\%  

& 0  & \cellcolor[rgb]{ .930,  .930,  .930}0\%    & 0 & \cellcolor[rgb]{ .930,  .930,  .930}0\%  & 0 & \cellcolor[rgb]{ .930,  .930,  .930}0\%

& 6 & \cellcolor[rgb]{ .930,  .930,  .930}5.9\%   & 6 & \cellcolor[rgb]{ .930,  .930,  .930}5.9\% & 0 & \cellcolor[rgb]{ .930,  .930,  .930}0\% \\
& GPT4-multi      
& 8 & \cellcolor[rgb]{ .930,  .930,  .930}13.1\%  & 8 & \cellcolor[rgb]{ .930,  .930,  .930}13.1\%    & 0  & \cellcolor[rgb]{ .930,  .930,  .930}0\%   
& 0 & \cellcolor[rgb]{ .930,  .930,  .930}0\%    & 0 & \cellcolor[rgb]{ .930,  .930,  .930}0\%    
& 0  & \cellcolor[rgb]{ .930,  .930,  .930}0\% 

& 8 & \cellcolor[rgb]{ .930,  .930,  .930}7.8\%   & 8 & \cellcolor[rgb]{ .930,  .930,  .930}7.8\% & 0  & \cellcolor[rgb]{ .930,  .930,  .930}0\% \\
& \textbf{\tool}   
& \textbf{29} & \cellcolor[rgb]{ .930,  .930,  .930}\textbf{47.5\%}  & \textbf{19} & \cellcolor[rgb]{ .930,  .930,  .930}\textbf{31.1\%}  & \textbf{4} & \cellcolor[rgb]{ .930,  .930,  .930}\textbf{6.6\%}
& \textbf{10} & \cellcolor[rgb]{ .930,  .930,  .930}\textbf{24.4\%} & \textbf{7} & \cellcolor[rgb]{ .930,  .930,  .930}\textbf{17.1\%} & \textbf{4} & \cellcolor[rgb]{ .930,  .930,  .930}\textbf{9.8\%}
& \textbf{39} & \cellcolor[rgb]{ .930,  .930,  .930}\textbf{38.2\%} & \textbf{26} & \cellcolor[rgb]{ .930,  .930,  .930}\textbf{25.5\%} & \textbf{8} & \cellcolor[rgb]{ .930,  .930,  .930}\textbf{7.8\%}\\ \midrule
% \multirow{3}{*}{Top-3}
\multirow{3}{*}{\begin{sideways}Top-3\end{sideways}} 
& GPT-4         
& 6 & \cellcolor[rgb]{ .930,  .930,  .930}9.8\%  & 6 & \cellcolor[rgb]{ .930,  .930,  .930}9.8\% & 1 & \cellcolor[rgb]{ .930,  .930,  .930}1.6\% 
&0 & \cellcolor[rgb]{ .930,  .930,  .930}0\%    & 0 &\cellcolor[rgb]{ .930,  .930,  .930}0\% & 0 &\cellcolor[rgb]{ .930,  .930,  .930}0\% 
& 6 & \cellcolor[rgb]{ .930,  .930,  .930}5.9\% & 6 & \cellcolor[rgb]{ .930,  .930,  .930}5.9\% & 1 &\cellcolor[rgb]{ .930,  .930,  .930}1.0\% \\
& GPT4-multi   
& 8 & \cellcolor[rgb]{ .930,  .930,  .930}13.1\%  & 8 & \cellcolor[rgb]{ .930,  .930,  .930}13.1\%   & 1 & \cellcolor[rgb]{ .930,  .930,  .930}1.6\%
& 0 & \cellcolor[rgb]{ .930,  .930,  .930}0\%    & 0 & \cellcolor[rgb]{ .930,  .930,  .930}0\% & 0 &\cellcolor[rgb]{ .930,  .930,  .930}0\% 
& 8 & \cellcolor[rgb]{ .930,  .930,  .930}7.8\% & 8 & \cellcolor[rgb]{ .930,  .930,  .930}7.8\% & 1 &\cellcolor[rgb]{ .930,  .930,  .930}1.0\%\\
& \textbf{\tool} 
& \textbf{32} & \cellcolor[rgb]{ .930,  .930,  .930}\textbf{52.5\%} & \textbf{23} & \cellcolor[rgb]{ .930,  .930,  .930}\textbf{37.7\%} & \textbf{12} & \cellcolor[rgb]{ .930,  .930,  .930}\textbf{19.7\%} 
& \textbf{13} & \cellcolor[rgb]{ .930,  .930,  .930}\textbf{31.7\%} & \textbf{10} & \cellcolor[rgb]{ .930,  .930,  .930}\textbf{24.4\%} & \textbf{7} & \cellcolor[rgb]{ .930,  .930,  .930}\textbf{17.1\%} 
& \textbf{45} & \cellcolor[rgb]{ .930,  .930,  .930}\textbf{41.1\%} & \textbf{33} & \cellcolor[rgb]{ .930,  .930,  .930}\textbf{32.4\%} & \textbf{19} & \cellcolor[rgb]{ .930,  .930,  .930}\textbf{18.6\%} \\ \midrule
% \multirow{3}{*}{Top-5}
\multirow{3}{*}{\begin{sideways}Top-5\end{sideways}} 
& GPT-4         
& 7 & \cellcolor[rgb]{ .930,  .930,  .930}11.5\%  & 7 & \cellcolor[rgb]{ .930,  .930,  .930}11.5\%  & 1 & \cellcolor[rgb]{ .930,  .930,  .930}1.6\%  
& 0  & \cellcolor[rgb]{ .930,  .930,  .930}0\%    & 0 & \cellcolor[rgb]{ .930,  .930,  .930}0\% & 0 & \cellcolor[rgb]{ .930,  .930,  .930}0\% 
& 7 & \cellcolor[rgb]{ .930,  .930,  .930}6.9\% & 7 & \cellcolor[rgb]{ .930,  .930,  .930}6.9\% & 1 & \cellcolor[rgb]{ .930,  .930,  .930}1.0\% \\
& GPT4-multi   
& 9 & \cellcolor[rgb]{ .930,  .930,  .930}14.8\%  & 9 & \cellcolor[rgb]{ .930,  .930,  .930}14.8\%   & 1 & \cellcolor[rgb]{ .930,  .930,  .930}1.6\% 
& 0 & \cellcolor[rgb]{ .930,  .930,  .930}0\%    & 0 & \cellcolor[rgb]{ .930,  .930,  .930}0\% & 0 & \cellcolor[rgb]{ .930,  .930,  .930}0\% 
& 9 & \cellcolor[rgb]{ .930,  .930,  .930}8.8\% & 9 & \cellcolor[rgb]{ .930,  .930,  .930}8.8\% & 1 & \cellcolor[rgb]{ .930,  .930,  .930}1.0\% \\
& \textbf{\tool} 
& \textbf{35} & \cellcolor[rgb]{ .930,  .930,  .930}\textbf{57.4\%} & \textbf{25} & \cellcolor[rgb]{ .930,  .930,  .930}\textbf{41.0\%} & \textbf{12} & \cellcolor[rgb]{ .930,  .930,  .930}\textbf{19.7\%} 
& \textbf{14} & \cellcolor[rgb]{ .930,  .930,  .930}\textbf{34.1\%} & \textbf{10} & \cellcolor[rgb]{ .930,  .930,  .930}\textbf{24.4\%} & \textbf{7} & \cellcolor[rgb]{ .930,  .930,  .930}\textbf{17.1\%} 
& \textbf{49} & \cellcolor[rgb]{ .930,  .930,  .930}\textbf{48.0\%} & \textbf{35} & \cellcolor[rgb]{ .930,  .930,  .930}\textbf{34.3\%} & \textbf{19} & \cellcolor[rgb]{ .930,  .930,  .930}\textbf{18.6\%} \\ 
\bottomrule
\end{tabular}}
% \vspace{-10pt}
\end{table}

% \liu{@yb:Add the description of Table VII here.}
As shown in Table~\ref{tab:rq2.2}, even with perfect location information, \tool consistently outperforms GPT-4. 
This indicates that \tool's patch generation capability is superior to that of GPT-4, even after multiple iterations. 
Compared with the results in Table~\ref{tab:rq2.1}, \tool only slightly improves its fixing effectiveness with perfect location information. While it does generate new plausible or correct patches, it sometimes fails to recreate some of its previous patches without perfect locations. 
This is because some patches were applied at related but different locations, such as within a method call hierarchy, and still effectively fixed the issue.
However, the provided location may interfere with the selection of fix patterns.
As for ChatGPT, when provided with exact fault locations, it can generate a few effective patches for large datasets, a stark improvement over its complete failure without precise locations. This further underscores the deficiency in GPT-4's fault localization capabilities.

\begin{table}[htbp]
\centering
\footnotesize
\caption{The number of panic-eliminated/plausible/correct patches of different tools on \dataset with perfect locations.} 
\label{tab:perfect}
\vspace{-5pt}
\label{tab:rq2.2}
\setlength{\tabcolsep}{0.6mm}{ 
\begin{tabular}{cc|cc|cc|cc|cc|cc|cc||cc|cc|cc}
\toprule
\multicolumn{2}{c|}{\multirow{2}{*}{\textbf{Tool}}} & \multicolumn{6}{c|}{\textbf{\dataset-Small (61)}}                                         & \multicolumn{6}{c||}{\textbf{\dataset-Large (41)}}          & \multicolumn{6}{c}{\textbf{Total (102)}}                \\ %\cline{3-8} 
&   & \multicolumn{2}{c|}{\textbf{Eliminated}} & \multicolumn{2}{c|}{\textbf{Plausible}}      & \multicolumn{2}{c|}{\textbf{Correct}}       & \multicolumn{2}{c|}{\textbf{Eliminated}} & \multicolumn{2}{c|}{\textbf{Plausible}}   & \multicolumn{2}{c||}{\textbf{Correct}}  & \multicolumn{2}{c|}{\textbf{Eliminated}} & \multicolumn{2}{c|}{\textbf{Plausible}}  & \multicolumn{2}{c}{\textbf{Correct}} \\ \midrule
%\multirow{3}{*} {Top-1}
\multirow{3}{*}{\begin{sideways}Top-1\end{sideways}} 
& GPT-4          
& 6 & \cellcolor[rgb]{ .930,  .930,  .930}9.8\%  & 4 & \cellcolor[rgb]{ .930,  .930,  .930}6.6\% & 4 & \cellcolor[rgb]{ .930,  .930,  .930}6.6\%
& 3  & \cellcolor[rgb]{ .930,  .930,  .930}7.3\%    & 3 & \cellcolor[rgb]{ .930,  .930,  .930}7.3\%    & 3 & \cellcolor[rgb]{ .930,  .930,  .930}7.3\%    
& 9 & \cellcolor[rgb]{ .930,  .930,  .930}8.8\%   & 7 & \cellcolor[rgb]{ .930,  .930,  .930}6.9\% & 7 & \cellcolor[rgb]{ .930,  .930,  .930}6.9\% \\
& GPT4-multi      
& 6 & \cellcolor[rgb]{ .930,  .930,  .930}9.8\%  & 4 & \cellcolor[rgb]{ .930,  .930,  .930}6.6\% & 4 & \cellcolor[rgb]{ .930,  .930,  .930}6.6\%
& 3 & \cellcolor[rgb]{ .930,  .930,  .930}7.3\%    & 3 & \cellcolor[rgb]{ .930,  .930,  .930}7.3\%    & 3 & \cellcolor[rgb]{ .930,  .930,  .930}7.3\%    
& 9 & \cellcolor[rgb]{ .930,  .930,  .930}8.8\%   & 7 & \cellcolor[rgb]{ .930,  .930,  .930}6.9\% & 7 & \cellcolor[rgb]{ .930,  .930,  .930}6.9\% \\
& \textbf{\tool}   
& \textbf{33} & \cellcolor[rgb]{ .930,  .930,  .930}\textbf{54.1\%}  & \textbf{24} & \cellcolor[rgb]{ .930,  .930,  .930}\textbf{39.3\%} & 7 & \cellcolor[rgb]{ .930,  .930,  .930}11.5\%  
& \textbf{10} & \cellcolor[rgb]{ .930,  .930,  .930}\textbf{24.4\%} & \textbf{7} & \cellcolor[rgb]{ .930,  .930,  .930}\textbf{17.1\%} & 4 & \cellcolor[rgb]{ .930,  .930,  .930}9.8\% 
& \textbf{43} & \cellcolor[rgb]{ .930,  .930,  .930}\textbf{42.2\%} & \textbf{31} & \cellcolor[rgb]{ .930,  .930,  .930}\textbf{30.4\%} & 11 & \cellcolor[rgb]{ .930,  .930,  .930}10.8\% \\ \midrule
% \multirow{3}{*}{Top-3}
\multirow{3}{*}{\begin{sideways}Top-3\end{sideways}} 
& GPT-4         
& 6 & \cellcolor[rgb]{ .930,  .930,  .930}9.8\%  & 5 & \cellcolor[rgb]{ .930,  .930,  .930}8.2\% & 4 & \cellcolor[rgb]{ .930,  .930,  .930}6.6\% 
&3 & \cellcolor[rgb]{ .930,  .930,  .930}7.3\%    & 3 &\cellcolor[rgb]{ .930,  .930,  .930}7.3\%  & 3 &\cellcolor[rgb]{ .930,  .930,  .930}7.3\%
& 9 & \cellcolor[rgb]{ .930,  .930,  .930}8.8\% & 8 & \cellcolor[rgb]{ .930,  .930,  .930}7.8\%  & 7 &\cellcolor[rgb]{ .930,  .930,  .930}6.9\%\\
& GPT4-multi   
& 6 & \cellcolor[rgb]{ .930,  .930,  .930}9.8\%  & 5 & \cellcolor[rgb]{ .930,  .930,  .930}8.2\%  & 4 &\cellcolor[rgb]{ .930,  .930,  .930}6.6\%
& 3 & \cellcolor[rgb]{ .930,  .930,  .930}7.3\%    & 3 & \cellcolor[rgb]{ .930,  .930,  .930}7.3\%     & 3 &\cellcolor[rgb]{ .930,  .930,  .930}7.3\%
& 9 & \cellcolor[rgb]{ .930,  .930,  .930}8.8\% & 8 & \cellcolor[rgb]{ .930,  .930,  .930}7.8\%  & 7 &\cellcolor[rgb]{ .930,  .930,  .930}6.9\%\\
& \textbf{\tool} 
& \textbf{36} & \cellcolor[rgb]{ .930,  .930,  .930}\textbf{59.0\%} & \textbf{27} & \cellcolor[rgb]{ .930,  .930,  .930}\textbf{44.3\%}  & 15 &\cellcolor[rgb]{ .930,  .930,  .930}24.6\%
& \textbf{13} & \cellcolor[rgb]{ .930,  .930,  .930}\textbf{31.7\%} & \textbf{11} & \cellcolor[rgb]{ .930,  .930,  .930}\textbf{26.8\%}  & 9 &\cellcolor[rgb]{ .930,  .930,  .930}22.0\%
& \textbf{49} & \cellcolor[rgb]{ .930,  .930,  .930}\textbf{48.0\%} & \textbf{38} & \cellcolor[rgb]{ .930,  .930,  .930}\textbf{37.3\%}  & 24 &\cellcolor[rgb]{ .930,  .930,  .930}23.5\%\\ \midrule
% \multirow{3}{*}{Top-5}
\multirow{3}{*}{\begin{sideways}Top-5\end{sideways}} 
& GPT-4         
& 6 & \cellcolor[rgb]{ .930,  .930,  .930}9.8\%  & 5 & \cellcolor[rgb]{ .930,  .930,  .930}8.2\% & 4 & \cellcolor[rgb]{ .930,  .930,  .930}6.6\% 
& 4  & \cellcolor[rgb]{ .930,  .930,  .930}9.8\%    & 4 & \cellcolor[rgb]{ .930,  .930,  .930}9.8\%    & 4 & \cellcolor[rgb]{ .930,  .930,  .930}9.8\%
& 10 & \cellcolor[rgb]{ .930,  .930,  .930}9.8\% & 9 & \cellcolor[rgb]{ .930,  .930,  .930}8.8\% & 8 & \cellcolor[rgb]{ .930,  .930,  .930}7.8\%\\
& GPT4-multi   
& 6 & \cellcolor[rgb]{ .930,  .930,  .930}9.8\%  & 5 & \cellcolor[rgb]{ .930,  .930,  .930}8.2\% & 4 & \cellcolor[rgb]{ .930,  .930,  .930}6.6\%
& 4 & \cellcolor[rgb]{ .930,  .930,  .930}9.8\%    & 4 & \cellcolor[rgb]{ .930,  .930,  .930}9.8\%    & 4 & \cellcolor[rgb]{ .930,  .930,  .930}9.8\%
& 10 & \cellcolor[rgb]{ .930,  .930,  .930}9.8\% & 9 & \cellcolor[rgb]{ .930,  .930,  .930}8.8\% & 8 & \cellcolor[rgb]{ .930,  .930,  .930}7.8\%\\
& \textbf{\tool} 
& \textbf{36} & \cellcolor[rgb]{ .930,  .930,  .930}\textbf{59.0\%} & \textbf{27} & \cellcolor[rgb]{ .930,  .930,  .930}\textbf{44.3\%} & 15 &\cellcolor[rgb]{ .930,  .930,  .930}24.6\%
& \textbf{13} & \cellcolor[rgb]{ .930,  .930,  .930}\textbf{31.7\%} & \textbf{11} & \cellcolor[rgb]{ .930,  .930,  .930}\textbf{26.8\%} & 9 &\cellcolor[rgb]{ .930,  .930,  .930}22.0\%
& \textbf{49} & \cellcolor[rgb]{ .930,  .930,  .930}\textbf{48.0\%} & \textbf{38} & \cellcolor[rgb]{ .930,  .930,  .930}\textbf{37.3\%} & 24 &\cellcolor[rgb]{ .930,  .930,  .930}23.5\%\\ 
\bottomrule
\end{tabular}}
\vspace{-10pt}
\end{table}

Regarding to the panic-eliminated and plausible patches, we believe that, when coupled with the textual interpretations derived from our mined patterns, panic-eliminated patches can aid developers repair their program efficiently.
This is because the failure of some patches to pass the regression tests may be attributed to additional logic after the patch is applied, which is beyond APR tool's capabilities.
% For instance, one unsuccessful patch generated by \tool involved the addition of an \texttt{as\_bytes()} method call to a variable for its return value, mirroring the developers' own solution. 
% This modification necessitated altering the return type of the variable, resulting in issues not directly related to the original bug, a phenomenon identified in prior research~\cite{yoo2012regression, huang2023survey}. 
% Nonetheless, the patches for panic bugs provided by \tool are fundamentally correct.
Such panic-eliminated fixes, with minimal manual oversight, can be comprehensively resolved, thereby reducing the developers' burden in understanding panic bugs and evaluating alternative functions.

\subsubsection{Efficiency}
\begin{wrapfigure}{r}{6cm}%靠文字内容的右侧
\centering
% \vspace{-15pt}
  \includegraphics[width=0.95\linewidth]{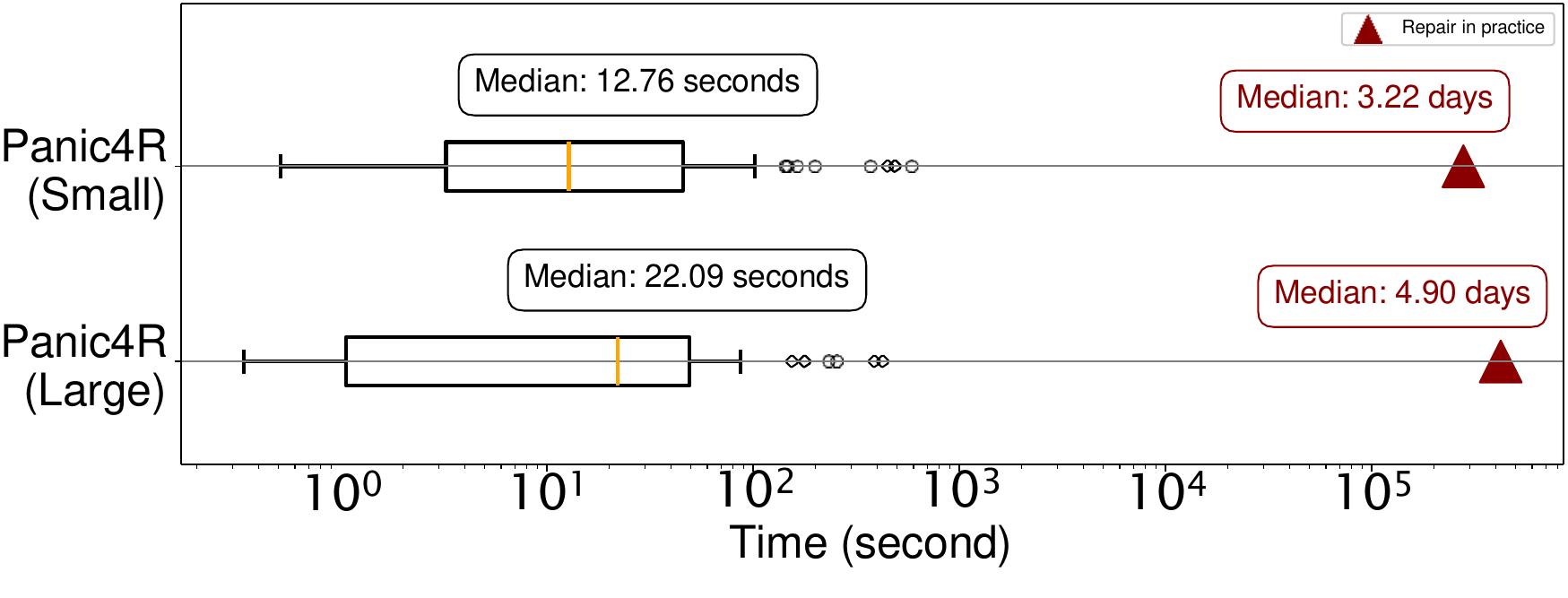}
  \vspace{-15pt}
  \caption{Experiment results of efficiency.}
  \vspace{-15pt}
  \label{fig:rq3}
\end{wrapfigure}

Fig~\ref{fig:rq3} and Table~\ref{tab:rq3} show the distribution of time spent on \tool and manual fixes.
On average, \tool requires about one minute to repair panics, whereas manual repairs can take days.
An interesting observation is that the average time of both \tool and manual repair are generally longer for smaller datasets. 
For \tool, this extended time is mainly due to the large test suites in some libraries, which lengthen the validation process. 
In terms of manual repairs, one notable case involved a panic bug that took over two years to fix, largely due to delayed maintenance by developers, resulting in a notably prolonged repair time for a smaller dataset.

% \begin{figure}[htbp]
%     \centering
%     \includegraphics[width=0.95\linewidth]{fig/rq3_boxplot.pdf}
%     \vspace{-10pt}
%     \caption{Experiment results of efficiency.}
%     \label{fig:rq3}
%     \vspace{-10pt}
% \end{figure}

In examining real-world PRs, we found a few key factors that potentially affect the bug-fixing time.
Developers often spend several days discussing issues due to the need to comprehend error messages, understand collaborators' code, and locate faults. 
Conversely, \tool provides detailed explanations with fix patches to streamline the review process and improve developer understanding. 
Additionally, maintainers sometimes request multiple revisions of a submitted fix, requiring different approaches and thus increasing time and effort. 
\tool addresses this by generating a range of patches, ranking them based on confidence scores and validation results, thus offering a prioritized patch list.

%%%%%%%%%%%%%%%%%%%%%%%%%%%%%%%%%
\begin{table}[htbp]
\centering
\footnotesize
\caption{Time consuming for \tool and actual fixes.} 
\vspace{-10pt}
\label{tab:rq3}
\setlength{\tabcolsep}{1.5mm}{ 
\begin{tabular}{ccccc}
\toprule
\textbf{Datasets}    & \textbf{Approaches}   & \textbf{Min} & \textbf{Max}     & \textbf{Avg.}  \\ \midrule
\textbf{\dataset-} & \tool & 0.51s   & 588.72s     & 65.77s  \\
\textbf{Small} & \cellcolor[rgb]{ .930,  .930,  .930}Actual     & \cellcolor[rgb]{ .930,  .930,  .930}0.52h    & \cellcolor[rgb]{ .930,  .930,  .930}807 days & \cellcolor[rgb]{ .930,  .930,  .930}76 days \\ \midrule
\textbf{\dataset-} & \tool & 0.34s  & 425.32s     & 61.29s      \\
\textbf{Large} & \cellcolor[rgb]{ .930,  .930,  .930}Actual      & \cellcolor[rgb]{ .930,  .930,  .930}0.42h   & \cellcolor[rgb]{ .930,  .930,  .930}236 days   & \cellcolor[rgb]{ .930,  .930,  .930}44 days \\
\bottomrule
\end{tabular}}
\vspace{-10pt}
\end{table}

\subsubsection{Case Study}
The overall evaluation results on real-world projects are presented in Table~\ref{tab:rq4}, showing that \tool successfully resolved 28 of 41 open issues with varied root causes, with all patches merged by developers. 
% These causes encompass arithmetic overflow, index-out-of-bounds errors, and unwrapping on None or invalid values, among others. 
% Besides, \tool has also generated some plausible patches that eliminated the panic, necessitating manual inspection and slight adjustments. 
As for unsuccessful fixes, some patches were plausible; they removed the panic but required further manual inspection and minor adjustments. 
% For some issues, no patch was generated due to the complexity of the panic's causes, suggesting the need to combine multiple repair patterns.
In terms of closed issues, \tool successfully generates 9 out of 22 patches that are similar to official patches.
The proportion is not very high, largely due to differences in their repair locations. For example, while the official approach might add a branch before the parameter call, \tool tends to insert it after creating the parameter. Nonetheless, both strategies achieve equivalent outcomes.

\begin{table}[htbp]
\centering
\footnotesize
\caption{Distribution of panics in real-world crates and fixing effectiveness of \tool.}
\vspace{-10pt}
\label{tab:rq4}
\setlength{\tabcolsep}{1.8mm}
%\begin{threeparttable}
\begin{tabular}{ccc|cc||cc}
\toprule
\multirow{2}{*}{\textbf{Crates}} & \multirow{2}{*}{\textbf{Stars}} & \multirow{2}{*}{\textbf{LoC}}& \multicolumn{2}{c||}{\textbf{Open issues}} & \multicolumn{2}{c}{\textbf{Closed issues}} \\
&   &  & \textbf{Total} & \textbf{Confirmed}      & \textbf{Total} & \textbf{Similar}    \\ \midrule
\textbf{hifitime}   & 315 & 10,762         & 28                & 21  & 0               & -          \\
\textbf{unicode-segmentation}  & 556 & 7,345            & 11                & 7   & 1               & 0     \\
\textbf{ratatui} & 9,086 & 41,534             & 1                 & 0      & 18              & 7  \\
\textbf{fancy-regex}  & 409 & 5,557       & 1                 & 0      & 3               & 2  \\ \midrule
\multicolumn{3}{c|}{\textbf{Total}}         & \textbf{41}                & \textbf{28}  & \textbf{22}              & \textbf{9}  \\ \bottomrule
\end{tabular}
\end{table}

Below we illustrate some bug cases and their fixes.

\begin{figure*}[tp]
    \centering
    \includegraphics[width=0.97\linewidth]{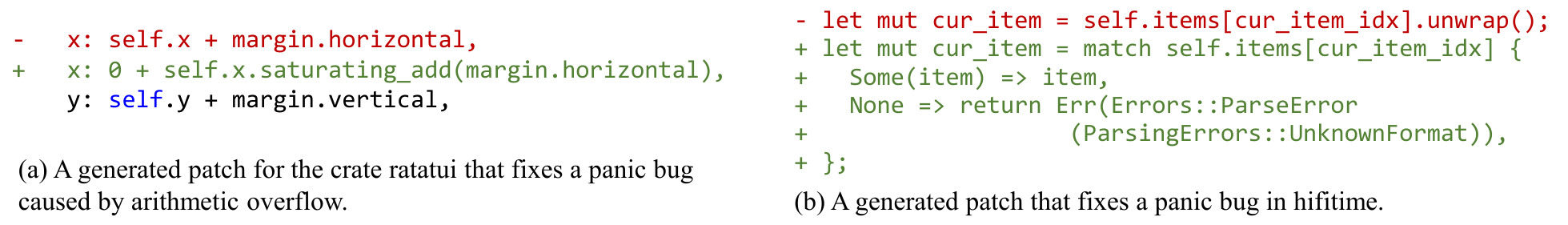}
    \vspace{-5pt}
    \caption{Patches generated by \tool have fixed real-world issues.}
    \label{fig:rq4}
    \vspace{-10pt}
\end{figure*}

\textbf{Bug Case1:}
Figure~\ref{fig:rq4} (a) shows the resolution of an arithmetic overflow error, which is a proposed patch of a closed issue for Rust crate ratatui~\cite{ratatui:online}.
Through fault localization, \tool succeeded in pinpointing a list of potential fault locations, prioritizing \texttt{src/layout.rs:233:20} as the top candidate, which aligns exactly with the actual fix.
Then, \tool successfully generated 3 patches for this location. 
In this case, the fault was identified within a struct filed expression; however, a closer inspection through iterative analysis revealed that this match expression included a binary expression, making it suitable for applying the \textit{Mutate Binary Operator} pattern. 
Finally, we calculate the similarity score between the generated patch and the official patch, obtaining a score of $0.95$, indicating the correctness. 

% \vspace{-10pt}
\textbf{Bug Case 2:}
The patch illustrated in Figure~\ref{fig:rq4} (b), generated by \tool, serves as an example of addressing a panic bug that arises from unwrapping a None/invalid value. 
This bug comes from an open issue of Rust crate hifitime~\cite{hifitime:online}, and our patch has been merged by developers.
\tool employs the \textit{Insert Match Unwrapper} pattern, specifically designed to mitigate problems stemming from the misuse of unwrap. 
This approach transitions the unsafe usage of \texttt{unwarp()} to pattern matching, which results in returning an error message rather than triggering a panic error. 
Additionally, the patch created by \tool takes into consideration the variable types to ensure consistency with the code context, thereby guaranteeing that the modified program passes regression testing.

% \section{Result Analysis}
% \input{sec/result} 

\section{Discussion}
This section discusses our infrastructure's application scenarios, comparison with existing works, and threats to validity.

\subsection{Application Scenarios}

This work establishes a systematic infrastructure that offers a diverse dataset and comprehensive analysis of panic bugs specific to Rust projects.
We now briefly discuss the application scenarios and the potential future research directions.

%\begin{itemize}
%[leftmargin=*]
\vspace{5pt}
\textbf{(1) Benchmark dataset for the Rust APR tools.}
The widely-used benchmark Defects4J has been crucial and influential for APR research in Java~\cite{liu2019avatar, koyuncu2019ifixr, liu2019tbar, xia2022alapha, wen2018capgen, chen2017jaid, chen2021jaid,  wong2021varfix}. 
Given the current limitations in the infrastructure for Rust program research, the dataset we propose, \dataset, is designed to serve as a robust and reliable benchmark for subsequent studies.
It includes macro switches and test case execution scripts, facilitating researchers to employ it as a benchmark dataset in their evaluations. 
    
% Inspired by the Defects4J construction methodology, \dataset is poised to serve as a similarly valuable benchmark for advancing APR research in the Rust community. 
\dataset offers detailed information for each case, encompassing the bug-triggering scenario and the exact patch applied to address the specific bug. This setup facilitates easy reproduction of the bug. 
Thus, \dataset enables further analysis to uncover underlying patterns and common root causes of panic bugs. This can lead to better insights into Rust's common vulnerabilities and areas where APR tools can focus or improve, ultimately contributing to the robustness of Rust codebases and reducing the likelihood of runtime panics in production software.
Moreover, it establishes a uniform evaluation metric, allowing for more objective and comparable assessments of APR tool effectiveness. Researchers and practitioners can measure improvements in patch accuracy, bug-fix efficiency, and detection of bug patterns across different tools, fostering a clearer understanding of APR advancements in the Rust ecosystem.

\vspace{5pt}
\textbf{(2) Extendable dataset for Rust program repair.}
Rust, as an emerging language with significant advantages in memory safety, increasingly research and approaches designed for Rust have been proposed. 
In recent years, some of the existing tools~\cite{deligiannis2023rustassistant, yang2024lancet, robati2024patterns} have been developed for fixing unique bugs in Rust programs. 
However, due to the less mature infrastructure of Rust compared to other mainstream languages, there has been a lack of structured and open-source datasets. 
This gap has made it challenging to standardize testing, repairing, and verification processes, and the collected datasets and code snippets are often not reusable for subsequent research. 

In our proposed \infra, the project structure and test scripts included in \dataset are both unified and highly extensible, including executable code snippets, test cases, and validation scripts. 
Future studies that aim to incorporate new panic bug repair datasets can easily extend \dataset using a standardized format. Similarly, for other types of bug repair datasets, maintaining a similar repository structure allows for quick adaptation to standardized dataset formats, minimizing redundant and tedious work.

% The project structure and test scripts included in \dataset are both unified and highly extensible. Researchers interested in Rust and program repair can leverage this foundational dataset to expand into other bug types or develop new repair patches.

\vspace{5pt}
\textbf{(3) Templates for pattern-based APR tools development.}
Pattern-based (also known as "template-based") APR tools was proved to be the most effective method compared to other approaches, such as search-based and constraint-based schemes~\cite{huang2023survey}.
After fault localization, the pattern-based APR tool could target these defects to select the corresponding fix templates to generate candidate patches.
The mined patterns in our proposed \infra, containing 34 sub-patterns addressing 9 types of panic root causes, can serve as templates for pattern-based APR tools. 
Based on this foundation, researchers can concentrate on developing improved techniques for fault localization and pattern prioritization.
For example, multiple potential patterns may match for the identified fault locations, necessitating further research into efficient pattern selection and validation. Additionally, ensuring the correctness of program semantics following the application of these patterns merits deeper investigation. 
    
In this paper, we demonstrate the effectiveness and practicality of these mined patterns through the implementation of \tool. Future work could focus on developing more refined repair tools based on these patterns, aimed at enhancing the accuracy of program semantics and improving repair efficiency.
    % In this paper, we develop \tool as an application to prove the effectiveness of these patterns.
    
    % \item \textbf{Educational Resources for Rust learners.}
    % We have developed a website that details the code changes, underlying root causes, and real-world examples for each fix pattern. This platform could serve as an educational resource to address the limitation of Rust-related learning materials and assist developers in understanding common solutions for panic fixes, enabling them to rectify issues in their programs.

\vspace{5pt}
\textbf{(4) Dataset for fine-tuning large language models (LLMs).}
In recent years, several research have employed deep learning models for automated program repair~\cite{xia2022alapha, fu2022vulrepair, chi2022seqtransautomaticvulnerabilityfix}, and their results have demonstrated the feasibility of using AI technologies to assist in fixing programs. 
However, the effectiveness of these approaches is often constrained by several factors, such as the scale of the datasets for model training, the risk of overfitting due to model fine-tuning, and the loss of information associated with tokenizing unstructured data like code and text~\cite{zhang2023survey}. 
With the rapid advancement of large-scale models, it is highly potential that program repair techniques based on these LLMs will be implemented. This development could significantly enhance the precision and efficiency of tradition learning-based automated repairs, mitigating current limitations and opening new avenues for research in software maintenance.

To support research in LLM-based program repair, high-quality datasets are essential~\cite{gupta2017deepfix, berabi2021tfix}. The datasets and mined patterns included in our proposed \infra, are particularly well-suited for use as training data for LLMs. 
Notably, the patterns we have identified are diverse, encompassing 34 sub-patterns and covering 9 different root causes of panic errors. Moreover, our open-sourced patterns feature various data structures, including source code, abstract AST structures, and corresponding textual descriptions. These rich sources of information are highly beneficial for the learning and evolution of LLMs, providing a robust foundation for developing more effective automated program repair technologies.

% As LLMs rapidly evolve, using high-quality data for fine-tuning can enhance their domain-specific applications. 
% By leveraging our carefully curated \dataset, which includes natural language interpretations and code change examples, researchers could fine-tune LLMs to develop AI-driven techniques for panic bug repair.

\vspace{5pt}
\textbf{(5) Providing informative suggestions for users.}
Our observations during data collection reveal that Rust developers, particularly beginners, often experience a sense of concern towards panic bugs. 
Additionally, the complex stack traces provided by the compiler can be difficult to understand. 
If tools were available to assist programmers in comprehending the causes of panic bugs and offering relevant repair suggestions, it would significantly reduce the complexity of understanding the program. 

In our \infra, \tool produces a mix of suggestions—code patches and textual interpretations—that are more informative than compiler raw output.
Given the steep learning curve of programs like Rust, we believe providing enhanced feedback is beneficial for developers. Such tools could demystify the error handling process, enabling developers to more effectively address and rectify issues, thereby enhancing their confidence and proficiency in using Rust.

\vspace{5pt}
\textbf{(6) Helpful for fixing real-world opening panic bugs.}
% \tool is both practical and effective, offering significant improvements in efficiency and effectiveness for repairing panic bugs.
% As a foundational infrastructure, practicality is one of the most important metric. 
The infrastructure we propose, \infra, is closely aligned with real-world application scenarios. 
The \dataset is derived entirely from open-source Rust projects within the ecosystem, ensuring its relevance and applicability. 
The patterns we have mined originate from Rust's official code implementations, further validating the authenticity and utility of our infrastructure. 
Moreover, \tool is capable of handling real, large-scale Rust projects. 
Our experiment results have shown \tool is much more efficient than manual fixes, and \tool has successfully resolved 28 panic bugs in real Rust crates, demonstrating its practical effectiveness.

%\end{itemize}

\vspace{-5pt}
\subsection{Comparison with Existing Code Patterns}

\textbf{Comparison with Java/C/C++ code patterns.}
Although numerous bug-fixing patterns oriented towards Java, C, and C++ have been proposed to corresponding APR tools, the code patterns we have identified are specific to Rust and are utilized for fixing panic bugs.
% Compared to fix patterns in Java or C++, our proposed patterns are more closely related to Rust-specific features. 
Specifically, some of the patterns, e.g., \textit{Reorder State Changer}, are designed to handle concurrency panics unique to Rust.
The patterns such as \textit{Insert Unsafe Block} are employed to address the unique unsafe features in Rust. 
% Additionally, our patterns are designed to handle concurrency panics unique to Rust. 
Additionally, unlike Java or C++, where safety rules such as ownership are not enforced by the language, our proposed patterns carefully maintain these safety rules. 
This attention to Rust's unique ownership model ensures that the fixes not only resolve the bugs but also uphold the language's guarantees of memory safety and concurrency. 
% Consequently, our approach is tailored to preserve Rust's stringent safety guarantees while addressing its distinct runtime challenges.

\vspace{5pt}
\textbf{Comparison with Rust-specific code patterns.}
To date, only a few fix patterns have been specifically developed for Rust programs. 
Rust-lancet~\cite{yang2024lancet} and Ruxanne~\cite{robati2024patterns} are the primary studies that have designed bug and fix code patterns uniquely for Rust. 
Unlike these studies, which focus on ownership-related or other common bug types such as missing attributes, our infrastructure addresses the most critical panic bugs in Rust. 
While bugs violating lifetime rules and other common issues usually prevent compilation, offering error messages and corrective suggestions, panic bugs manifest at runtime, causing abrupt program termination. 
These bugs, influenced by Rust’s distinct memory and process management, render existing fix patterns inadequate for addressing panic issues. 
This paper introduces the first infrastructure aimed at exploring Rust panic bugs, identifying their root causes and developing specific repair patterns, significantly enhancing the understanding and remediation of Rust programs.

\vspace{-6pt}
\subsection{Threats to Validity}
One of the potential threats to validity concerns the representativeness of our collected dataset, since all code and patches are sourced from open-source crates. 
However, we consider these crates to be relatively complex and representative of large-scale Rust projects. 
\dataset comprises the top 500 most downloaded crates, reflecting the actual usage frequency and activity levels of these programs. 
We have also modeled our data collection process after the Defects4J dataset to enhance the reliability of our data. 
This adherence to proven methodologies in dataset construction supports the validity of our research findings.

Another potential threats lies in the incompleteness of the mined fixing patterns. 
In real-world scenarios, specific bug triggers may have unique or diverse repair strategies. 
To enhance the comprehensiveness of our mined patterns, we systematically explore and extract from Rust's official implementation code, which is considered to contain the most standard repair strategies. 
Additional patterns that may emerge in the future can also be easily integrated into our released \infra, further improving the effectiveness of pattern-based repair tools.
 
% \vspace{-4pt}
\section{Related Work}
% \vspace{-4pt}
In this section, we compare our work with other APR approaches, as well as the testing and analysing work for Rust program.

\textbf{\textit{Automated Program Repair.}}
Automated Program Repair (APR) has witnessed significant advancements in recent years. 
Most of methods~\cite{liu2019avatar, koyuncu2019ifixr, xia2022alapha, wen2018capgen, chen2017jaid, chen2021jaid, saha2019herclues, wong2021varfix, liu2019kpar, liu2019tbar} are designed for Java programs, with evaluated on Defect4J~\cite{just2014defect4j}, a dataset of real bugs from open source Java programs.
%These techniques can be broadly categorized into non-learning-based and learning-based approaches.
A recent study~\cite{huang2023survey} divides non-learning-based APR into three categories: search-based~\cite{weimer2013ae, qi2013trpautorepair, wen2018capgen, sidiroglou2015cp, le2012genprog}, constraint-based~\cite{pei2011autofix, pei2014autofix, wei2010autofix, chen2017jaid, chen2021jaid, afzal2021sosrepair, mechtaev2015directfix} and template-based~\cite{tan2015relifix, le2016hdrepair, saha2019herclues}. 
VarFix~\cite{wong2021varfix}, a search-based way of observing which combinations of edit operations pass the test. Constraint-based techniques like Nopol~\cite{xuan2017nopol} and SemFix~\cite{nguyen2013semfix} transform the repair process into a constraint solving problem, reducing the search space. 
kPAR~\cite{liu2019kpar} and AVATAR~\cite{liu2019avatar} generate fix patterns collected by manual extraction and static violation analysis respectively, used by iFixR~\cite{koyuncu2019ifixr}. 
TBar~\cite{liu2019tbar} is proposed to assess the qualitative and quantitative diversity of previous repair templates.
As for learning-based APR, AlaphaRepair~\cite{xia2022alapha} achieves state-of-the-art results on both Java and Python programs via zero-shot learning. 
VulRepair~\cite{fu2022vulrepair} highlights the advancement of NMT-based automated vulnerability repairs with pre-trained models.
% According to a survey~\cite{zhang2023survey} on learning-based APR, extensive manual effort is required to produce high-quality test datasets to train a reliable model, which implies the high cost in learning-based APR techniques. 
% Also, it has been questioned about the generalizability of learning-based APR tools when it comes to the strict syntactic structural features and complex semantic dependencies of PLs~\cite{zhang2023survey, jiang2021cure}, which our experiments with GPT-4 has demonstrated to some extent.

Different from existing APR tools that focus on Java and C programs~\cite{huang2023survey}, we represent the first dedicated infrastructure \infra targeted at Rust panic bugs.
Several tools~\cite{yang2024lancet, robati2024patterns, deligiannis2023rustassistant} have been proposed to address Rust compilation bugs. However, they are not specifically designed to target panic bugs that occur during runtime.
Considering the significant impact of Defects4J and the steep learning curve associated with Rust, we believe \infra would serve as a foundational resource for Rust's research. 
% It aims to provide a reliable infrastructure for further studies in Rust program comprehension, automated repair, and related areas.

% Defect4j~\cite{just2014defect4j}, a database of existing real bugs from open source Java programs, has always been served as the most widely used current benchmark~\cite{huang2023survey} for many APR tools~\cite{liu2019avatar, koyuncu2019ifixr, xia2022alapha, wen2018capgen, chen2017jaid, chen2021jaid, saha2019herclues, wong2021varfix, liu2019kpar, liu2019tbar}. 
% However, such benchmark for Rust panic bugs has never been proposed before and our database fills this gap by collecting real panic bugs from open-source Rust crates, which not only provides a valuable resource for understanding the nature and prevalence of Rust panic bugs but also guides the direction of research and development in the specific domain of Rust bug repair. 

\textbf{\textit{Rust Program Testing and Analysing.}}
Due to Rust's innovative safety mechanism, new challenges have been posed in its testing and analyzing. 
RustSmith~\cite{sharma2023rustsmith} employs random program generation to test the Rust compiler. 
As for RULF~\cite{jiang2022rulf} and SyRust~\cite{takashima2021syrust}, they concentrate on testing Rust crates by generating API sequences.
In the realm of Rust program analysis, RUPTA~\cite{li2024rupta} introduces a context-sensitive pointer analysis framework for Rust, successfully applied to construct call graphs. Additionally, through static analysis, tools like SafeDrop~\cite{cui2023safedrop} and Rudra~\cite{bae2021rudra} detect memory safety issues in large-scale Rust programs, contributing to enhanced program robustness. In contrast, RustCheck~\cite{xia2023rustcheck} employs dynamic analysis techniques to uncover memory safety vulnerabilities. 

Different from existing testing approaches, our work proposed the first APR tool specifically tailored to address errors related to Rust's panic mechanism, and we focused on fixing the practical panic bugs. 
Besides, we have constructed a real-world code dataset and fix patterns, which serves as an infrastructure for Rust program comprehension and repair.

\vspace{-10pt}
\section{Conclusion}
In this paper, we introduce an infrastructure \infra designed to support fixing panic bugs of real-world Rust programs.
We construct the first Rust program's fixing dataset, \dataset, containing $102$ real-world panic bugs and their patches.
Additionally, we conduct pattern mining based on Rust compiler source code to identify Rust-specific fixing patterns. 
We also introduce an APR tool, \tool, which effectively localizes faults and generates patches, outperforming commercial LLM-based tools. 
Moreover, \tool has successfully resolved 28 open issues related to panics, all of which have been confirmed and merged by developers.

\vspace{-10pt}
\section*{Data Availability}
The dataset, mined patterns, the source code of \tool, and experiment results can be found at: \url{https://sites.google.com/view/panickiller/home}.

% \medskip
% \noindent \textbf{Data Availability.} \dataset containing test cases and patches, as well as the source code of our tool can be accessed via this link: \url{https://anonymous.4open.science/r/PanicKiller-2024-828E/README.md}.

\bibliographystyle{ACM-Reference-Format}
\bibliography{arxiv}

\end{document}